# Topological Kinetic Crossover in a Nanomagnet Array

Xiaoyu Zhang[1], Grant Fitez[1], Shayaan Subzwari[1], Nicholas S. Bingham[1†], Ioan-Augustin Chioar[1], Hilal Saglam[1], Justin Ramberger[3], Chris Leighton[3], Cristiano Nisoli*[2], & Peter Schiffer*[1,]

[1]Department of Applied Physics, Yale University, New Haven, CT 06511, USA

[2]Theoretical Division and Center for Nonlinear Studies, MS B258, Los Alamos National Laboratory, Los Alamos, New Mexico 87545, USA

[3]Department of Chemical Engineering and Materials Science, University of Minnesota, Minneapolis, Minnesota 55455, USA

[4]Department of Physics, Yale University, New Haven, CT 06511, USA

**ABSTRACT**

**Ergodic kinetics, which are critical to equilibrium thermodynamics, can be constrained by a system's topology. We study a model nanomagnetic array in which such constraints visibly affect the behavior. In this system, magnetic excitations connect into thermally active one-dimensional strings whose motion can be imaged in real time. At high temperatures, we observe the merging, breaking, and reconnecting of strings, resulting in the system transitioning between topologically distinct configurations. Below a crossover temperature, the string motion is dominated by simple changes in length and shape. In this low temperature regime, the system is energetically stable because of its inability to explore all possible topological configurations. This kinetic crossover suggests a generalizable conception of topologically broken ergodicity and limited equilibration.**

*Corresponding authors' emails: peter.schiffer@yale.edu and cristiano@lanl.gov

†Present address: Department of Physics, University of Maine, Orono, Maine 04469, USA

Our understanding of equilibrium thermodynamics relies on simplifying assumptions about kinetics, e.g., the ergodic hypothesis, which postulates that a system can explore all energetically-equivalent configurations (see, for example [1]). When kinetics are constrained [2–7], however, ergodicity can break down on relevant timescales, thus causing memory effects, glassiness, non-equilibration, or slow equilibration [8].

Topological constraints on kinetics are of particular interest in the context of the critical role of topology in much of modern physics, and they have been examined within the context of various mathematical models [2,4]. The constraints can arise from a partition of the accessible states (or phase space) into topologically distinct subsets, often called topological sectors [9,10]. Kinetics within a sector are topologically trivial, i.e., they do not alter the system's topology, contrasting with nontrivial kinetics that allow the system to cross through sectors and change the topology of the system. If the latter are constrained, then the system cannot explore all of its configurations, and should show deviations from reversible equilibrium thermodynamics. In experimental systems, topologically constrained kinetics was invoked early in the context of soft matter systems, e.g., in macromolecule elasticity, foams, cellular patterns, and the kinetics of continuous networks [2], and the broad topic is closely connected to fundamental questions around equilibration, ergodicity, and memory.

**Strings in Santa Fe ice**

We report here the direct visualization of topologically constrained kinetics in a designed nanomagnet array. Such arrays, known broadly as artificial spin ice systems, have displayed a range exotic collective phenomena and technological potential [11,12]. The particular array geometry we study is Santa Fe ice (SFI) [13,14]. The structure of SFI is shown in Fig. 1A and B, where each element is a single domain ferromagnetic island with the moment oriented along the long axis due to shape anisotropy. The islands are arranged in rectangular plaquettes, with pairs of interior plaquettes (shaded in the figure) surrounded by six peripheral plaquettes. Much of the physics of this system can be understood through the moment configurations at the vertices where islands converge in the lattice [13,14]. The SFI lattice geometry requires that some fraction of the vertices are not in the lowest local energy state, even in the collective ground state. These local excitations have previously been dubbed "unhappy" vertices [13,15] and are highlighted as circular red dots in Figs. 1C and D. SFI lends itself to the study of topological kinetics because, as we had previously shown [14], the lattice geometry forces unhappy vertices to form one-dimensional strings that can either end within interior plaquettes or form closed loops (illustrated below and described in detail in section S3 [16]). Fig. 1C and D, show these strings as olive lines connecting the unhappy vertices in two realizations of the highly degenerate disordered ground state of the system [13,14]. Such strings naturally have complex topological properties [17], as is readily apparent in a bowl of spaghetti or udon [18,19].

To probe the above physics, we studied samples composed of stadium-shaped permalloy islands with designed lateral dimensions of 470 × 170 nm$^2$, thickness of ~2.5 nm, and lattice spacing of $a$ = 600, 700, and 800 nm as defined in Fig. 1A. The specifics of sample fabrication and characterization are in section S1 [16] and have been described previously [14], where we also showed that the string population and length distributions are thermally activated at the highest experimentally accessible temperatures. We measured the moment configuration in our samples using x-ray magnetic circular dichroism photoemission electron microscopy (XMCD-PEEM) [16], which yields real-space images of the island magnetic moments (Fig. 1B). Upon increasing temperature, thermal excitations caused a fraction of the moments to visibly flip in orientation between images, with the rate of flipping increasing with increasing temperature, and we could therefore observe the thermal kinetics of this system. We show results for the $a$ = 600 nm sample below, where the inter-island interactions are strongest; the data are consistent for the other lattice spacings [16]. All data are derived from the same XMCD-PEEM images taken on the same samples that were analyzed in reference [14].

Because the energy cost of a string increases with its length, the disordered ground state corresponds to strings of three unhappy vertices, each on three-island vertices. Therefore, in the ground state, strings are constrained to connecting near-neighbor interior plaquettes, as shown in Fig. 1C and D. In excited states of the system, strings can be longer and can connect interior plaquettes that are not neighboring, or can form closed loops. In our experimental data, the system approached the disordered ground state at the lowest temperatures, with a residual population of excitations that increased in number for the larger lattice spacings [14].

**Types of string motion**

In considering the string kinetics, one must consider the possible motions of the strings within the geometrical constraints imposed by the SFI structure. For any string configuration, one can continuously deform the strings by bending, elongating, or shrinking them, while keeping the end points the same. Such deformations do not change the topology of the string configuration and thus keep the system within a single topological sector of available configurations, also known as a "homotopy class". In the case of SFI, a homotopy class is defined as the set of all string configurations connecting the same interior plaquettes, in which any member of the set can be transformed to any other member of the set through only continuous deformations. Because the SFI ground state can be realized with many different sets of string connections among the plaquettes, the low energy configurations of the system can be partitioned into different homotopy classes (see Fig. 1C and D).

By contrast, string configurations that cannot be obtained by continuous bending or deformation of the strings are in different homotopy classes. For example, if two string configurations have different interior plaquettes attached to each other, there is no continuous way to transform the strings from one configuration to another. Strings must be created, eliminated, combined, or cut, to transform between such configurations, thus breaking continuity and changing the topology.

In terms of energy, each homotopy class possesses an energy minimum, obtained by continuously deforming and shrinking all the strings in any configuration belonging to the class, so that strings have the minimum number of lowest-energy vertices connecting with the string end points. If strings in a homotopy class extend to connect interior plaquettes that are not in close proximity, the energy minimum of that homotopy class is necessarily higher than the global ground state energy. To relax to the ground state, the system requires a topological change in the strings to change the homotopy class. Such a process, in which a string changes in such a way that the system transits from one homotopy class to another, is called "topological surgery" [20,21]. As discussed below, the specific process of topological surgery in SFI occurs when two or more strings merge and then separate into a new configuration. This concept of topological surgery, i.e., the connection and separation of proximate one-dimensional objects, can also be used to describe diverse phenomena such as chromosome meiosis, DNA recombination, magnetic flux reconnection in plasmas, vortex line fusion and reconnection in classical and quantum liquids, and dislocation lines in metallurgy [22–26]. SFI provides an unusual opportunity for experimentally tracking this process in real time in a thermal system.

Based on the above discussion, we now consider the kinetics of our string configurations by comparing sequential XMCD-PEEM images [27,28] and recording the motions (where we use the term "motion" to describe any change to a string, including the creation and annihilation of loops). We classify string motions into two categories: trivial and nontrivial. These two categories correspond to motions that continuously deform a string, or to motions that include a change in string topology, respectively. In other words, trivial motions do not change the homotopy class of the string configuration in the system, whereas nontrivial changes do. The distinction is demonstrated in Fig. 2, where we show simple examples of each, and in Fig. 3, where we give a taxonomy of trivial and nontrivial motions.

The "trivial motions" of strings correspond to continuous deformations of strings that retain the same interior plaquettes as endpoints. The different variations among these are shown in the top schematic of Fig. 3. As shown in the figure, the different types of trivial motions are labelled: 'wiggle' (red [T1], dark blue [T2], and purple [T3]); 'grow and shrink' (orange [T4]); 'loop wiggle' (green [T5]); 'loop grow and shrink' (light blue [T6]); and 'loop creation and annihilation' (pink [T7]). The loop motions are included as trivial because loops are contractible to zero, and thus loop creation and annihilation are topologically trivial.

The "nontrivial motions" of strings are those changes that represent a change in the homotopy class of the system, i.e., changing how the interior plaquette endpoints are connected by strings. The different variations among these are shown in the bottom schematic of Fig. 3. As shown in the figure, the different types of nontrivial motions are labelled: 'merge and split' (magenta/red [N1/N2] and light purple [N3]); 'reconnection' (green/light blue [N4/N5] and olive/dark blue [N6/N7]); 'adjacent reconnection' (purple [N8]); and 'strings on adjacent interior plaquettes creation and annihilation' (orange [N9]). Note that merge and split are necessarily intermediate steps of a reconnection, but because of the limited time resolution we label a full reconnection distinctly, when we can witness a full reconnection between two consecutive frames. Note also that adjacent reconnection and interior creation and annihilation are in a separate category in that they involve only adjacent interior plaquettes with a shared side and the flip of a single moment along that side. We therefore focus on the topologically nontrivial kinetics associated with strings that traverse the peripheral plaquettes.

In addition to the trivial and nontrivial string motions classified above, naturally, there are strings that do not change between sequential images, i.e., that have no motion. There are also 'complex' motions that involve multiple nontrivial changes in moment orientations between two sequential XMCD-PEEM images, as a consequence of the finite time resolution of image acquisition. Additionally, there are other trivial and nontrivial motions that are associated with a small number of nanoislands whose magnetizations could not be determined from the images (< 0.1%) or that lie on the edges of our XMCD images; both resulted in incomplete strings in our image recognition program. These other motions and complex motions are not listed in the table of Fig. 3.

**Quantifying string motion**

In order to quantitatively and unambiguously characterize the string motions from our sequential experimental maps of the moments, we used an analysis program that labels each string with two sets: one set is the interior plaquettes at which the string ends, and the other set is the unhappy vertices it connects. By comparing the elements in these two sets between sequential images, the algorithm can therefore collect the temperature dependent prevalence of each type of string motion [16]. We quantify the string motion prevalence as $N$, the average number of string motions between subsequent XMCD-PEEM images per unit cell. We note that $N$ effectively gives the rate of string motions, since the images are taken at one second intervals, and that it can be separated by the type of motion.

We now turn to the temperature dependence of our data. We first examine the total vertex energy of the system as a function of temperature, e.g., the sum of the magnetostatic energy at each vertex, averaged over all images at a given temperature, as determined by micromagnetic calculations [16,29]. As shown in Fig. 4A, the energy, quantified as the excess above the ground state, has a clear crossover around $T_X$ = 330 K, which is reflected in the kinetics of the strings as discussed below and is also visible in our previously published data on average string length in the system [14].

Fig. 4B-D show the statistics of motions as a function of temperature, with the temperature regime below $T_X$ indicated by shading. Below $T_X$, trivial string motions dominate, whereas nontrivial string motions dominate above $T_X$ (note that the number of trivial motions at higher temperatures is suppressed by a fraction of them being subsumed within the non-trivial motions). We note that the system energy in Fig. 4A is nearly flat below $T_X$, i.e., the system does not substantially change its energy upon cooling (even though string motions retain significant temperature dependence, as discussed below). This temperature independence is consistent with the observed crossover in the predominant types of string motions from nontrivial to trivial upon cooling through $T_X$. Significant reductions in system energy require topological changes in the strings, i.e., nontrivial motions, and therefore the system is unable to relax to a lower energy overall configuration.

Among the nontrivial motions above $T_X$, the "complex motions" are predominant (Fig. 4C). Because these represent multiple updates of the moments among PEEM images at higher temperature, they also include most other topologically nontrivial motions. By contrast, local processes among adjacent interior plaquettes dominate below $T_X$ (Fig. 4D). This is consistent with the system's topological constraints, as such motions require only a single moment flip, and they make a minimal change to the system's topology.

We now consider the trivial kinetics. In Fig. 5A we see that the loop processes (light green curves) and wiggles (red curves) are the highest contributors. The grow (blue) and shrink (violet) curves always follow each other, as expected from detailed balance. The same can be said in general for all reversible inelastic processes, as seen also in wiggles corresponding to energy increase and decrease in Fig. 5B (the increase in energy being caused by the strings occupying a vertex of coordination four, whose excitations have slightly higher energy than vertices of coordination 3) and loop creation and annihilation (Fig 5C). Most of the wiggles, however, do not change the total system energy (Fig. 5B), and thus they dominate (red curve in Fig. 5A).

Finally, in Fig. 5D we show a log linear plot of the number of trivial processes, $N_{trivial}$, versus the reciprocal temperature. Within our fairly narrow accessible temperature range below $T_X$, the data display apparent thermally activated behavior, i.e., $N_{trivial}(T) = \frac{1}{\tau_o} e^{-\frac{E_S}{T}}$, which is also observed for the other lattice spacings [16]. Fits of that form to the data, however, depend strongly on our choice of fitting range, with $\tau_o$ ~ $10^{-5}$ seconds and $E_S$ ~ 3000 K for the $a$ = 600 nm sample. We can qualitatively understand this behavior in that the motion of strings, even those with no net energy change, requires moments to flip in direction and overcome the energy barrier associated with the island's shape anisotropy. The energy barrier to flipping a moment, as calculated by micromagnetics at zero temperature, is considerably higher than $E_S$ [16], but we expect that the energy barrier for flipping the moment in a real system will be affected by the magnetic texture of the island at elevated temperature [30]. Similarly, the effective attempt frequency, $1/\tau_o$, is small compared to the typical spin wave frequency, which is in the GHz regime. Again, this is perhaps not surprising in that the string motions, even almost all the trivial ones, are associated with changes in multiple islands.

**Discussion and outlook**

Our observed crossover leads to a consistent picture of how the string kinetics evolve. Below $T_X$, the system does not change its energy, but it is kinetically active via topologically trivial motions (bending and stretching modes). In this regime, the system has a low likelihood to undergo topological surgery and change between homotopy classes. In other words, the limited ability of the system to explore homotopy classes likely prevents full ergodicity on experimental timescales at low temperatures, and thus the system is impeded from further reducing its global energy to fully realize the ground state. Note that the trivial kinetic processes are thermally equilibrated below $T_x$, but within a limited phase space of possible string configurations. This is evidenced by the close tracking of energy increasing and decreasing motions in Fig. 5A-C, which indicates that detailed balance holds, and the activated behavior in Fig. 5D.

At high temperature, above $T_X$, by contrast the topologically nontrivial motions become predominant and the system appears to be fully ergodic. This allows the total energy of the system to change significantly, because the different homotopy classes can have radically different energies, and this is the regime in which the activated string length was demonstrated previously [14].

The observation of a clear crossover between the two regimes correlates somewhat to glass transitions and other dynamic slowdowns, but here is driven by topology rather than a disordered potential landscape. We can broadly understand the crossover in that the homotopy-class-changing non-trivial motions require substantially longer strings and thus more moments to reverse directions than the trivial motions, and therefore are kinetically suppressed with decreasing temperature. While we do not associate the crossover with a thermodynamic phase transition and associated non-analytic behavior, the relative sharpness of the crossover is striking relative to most glass transitions, suggesting future studies with frequency-sensitive techniques [31–33].

Our findings should be generalizable in defining and characterizing kinetic crossovers to non-ergodicity beyond systems with topological constraints [34]. Comparing experimental results to existing theoretical methods to ascertain ergodicity could elucidate slow relaxation after quenches and memory effects and explore the relation with other kinetic crossovers [8,34–36]. Such studies could also have relevance to novel forms of computing [37], and possibly quantum tunneling through homotopy classes within qubit realizations [38] of SFI or similar structures. Our work might also prove pertinent to systems such as structural glasses, where dislocation lines ending in vertices proved an useful description [39–42]. The advantage of our physical realization of SFI is that it allows direct experimental measurements of these phenomena, suggesting future examinations of other bespoke topologically interesting structures.

**ACKNOWLEDGMENTS**

The authors thank Claudio Castelnovo, Nigel Goldenfeld and Yair Shokef for helpful feedback on the manuscript. We also thank Rajesh Chopdekar and Roland Koch for support at the PEEM-3 endstation at beamline 11.0.1.1 of the Advanced Light Source during our data acquisition.

**FUNDING:**

Work at Yale University was funded by the US Department of Energy, Office of Basic Energy Sciences, Materials Sciences and Engineering Division under Grant No. DE-SC0020162. This research used resources of the Advanced Light Source, a DOE Office of Science User Facility under contract no. DE-AC02-05CH11231. Work at the University of Minnesota was supported by NSF through Grant No. DMR-2103711. Work at Los Alamos National Laboratory was carried out under the auspices of the US Department of Energy through Los Alamos National Laboratory, operated by Triad National Security, LLC (Contract No. 892333218NCA000001) and financed by DoE LDRD.

**AUTHORS' CONTRIBUTIONS**

J. Ramberger performed film depositions under the guidance of C. Leighton. X. Zhang and N. S. Bingham oversaw the lithography. X. Zhang, N. S. Bingham, H. Saglam and I.-A. Chioar performed the PEEM characterization of the thermally active samples. X. Zhang, G. Fitez, and S. Subzwari analyzed the string structures. C. Nisoli developed the theory and wrote the first draft. P. Schiffer supervised the entire project. All authors contributed to the discussion of results and to the finalization of the manuscript.

**Competing interests:** None

**Data and materials availability:**

Additional experimental data and underlying data from the plots generated in this study are available at Dryad [42]

**Supplementary Materials**

Materials and methods
Supplementary text
Figs. S1-S6
Tables S1-S2

**Figure 1. The Santa Fe ice (SFI) geometry. (A)** Schematic of the lattice, where each element represents a single-domain nanomagnet, and the lattice spacing is *a*. The unit cell (indicated in yellow) has four interior plaquettes (indicated by light blue shading) that are separated by 2-island vertices, and twelve peripheral plaquettes that surround them. **(B)** X-ray magnetic circular dichroism photoemission electron microscopy (XMCD-PEEM) image of SFI, in which the entire islands are either black or white, indicating the magnetic moment direction through its component projected onto the incident X-ray beam direction (red arrow). Note that the lattice is slightly offset from the structure in (A) **(C, D)** Illustrations of two SFI disordered ground states, where the excited vertices are indicated by red circular dots. They are in different homotopy classes, as explained in the text, because the interior plaquettes that are connected by strings are distinct.

**Figure 2. Trivial and nontrivial string motions**. Red dots represent unhappy vertices. **(A-C)**. Examples of trivial string motions. Between configurations (A) and (B), there is a "wiggle" motion with no energy change. Between configurations (B) and (C) there is a "grow" motion with energy

change because (C) is longer. **(D-F)**. Examples of nontrivial motions. Configurations (D) and (F) have pairs of strings that have different pairs of end points, and configuration (E) has a merged crossed string with four end points. Each of the three configurations represents topological distinct states, i.e., different homotopy classes.

**Figure 3. Illustration and taxonomy of various string motions**. Island moments are shown as open arrows and the flipped island moments filled with color corresponding to the strings. The upper schematic shows trivial string motions. The lower schematic shows nontrivial string motions. The table lists the illustrated motions, as described in the text.

**Figure 4. Temperature-dependent string properties**. The shaded region indicates the low temperature regime below $T_X$. **(A)** The excess energy per unit cell versus temperature. **(B)** The temperature dependence of the average number of string motions per PEEM image. We also tracked the number of strings that did not change between subsequent images, and we plot the total number of such unchanging strings as 'no motion'. **(C)** The temperature dependence of the average number of string motions per PEEM image for all types of nontrivial motions. **(D)** An expanded plot of the low temperature regime from (C). All data are for $a$ = 600 nm SFI, and the error bars are the standard errors from all PEEM images taken at given temperature.

**Figure 5. Temperature dependence of trivial string motions.** Shown are the average number of string motions per PEEM image for trivial motions. The shading indicates the low temperature regime where trivial string motions dominate. **(A)** All types of trivial motions. **(B)** String wiggle subtypes of trivial motions. **(C)** Loop subtypes of trivial motions. **(D)** The number of total trivial motions as a function of inverse temperature with fit as described in the text. All data are shown per unit cell and for $a$ = 600 nm. The error bars are the standard errors from all XMCD-PEEM images taken at given temperature.

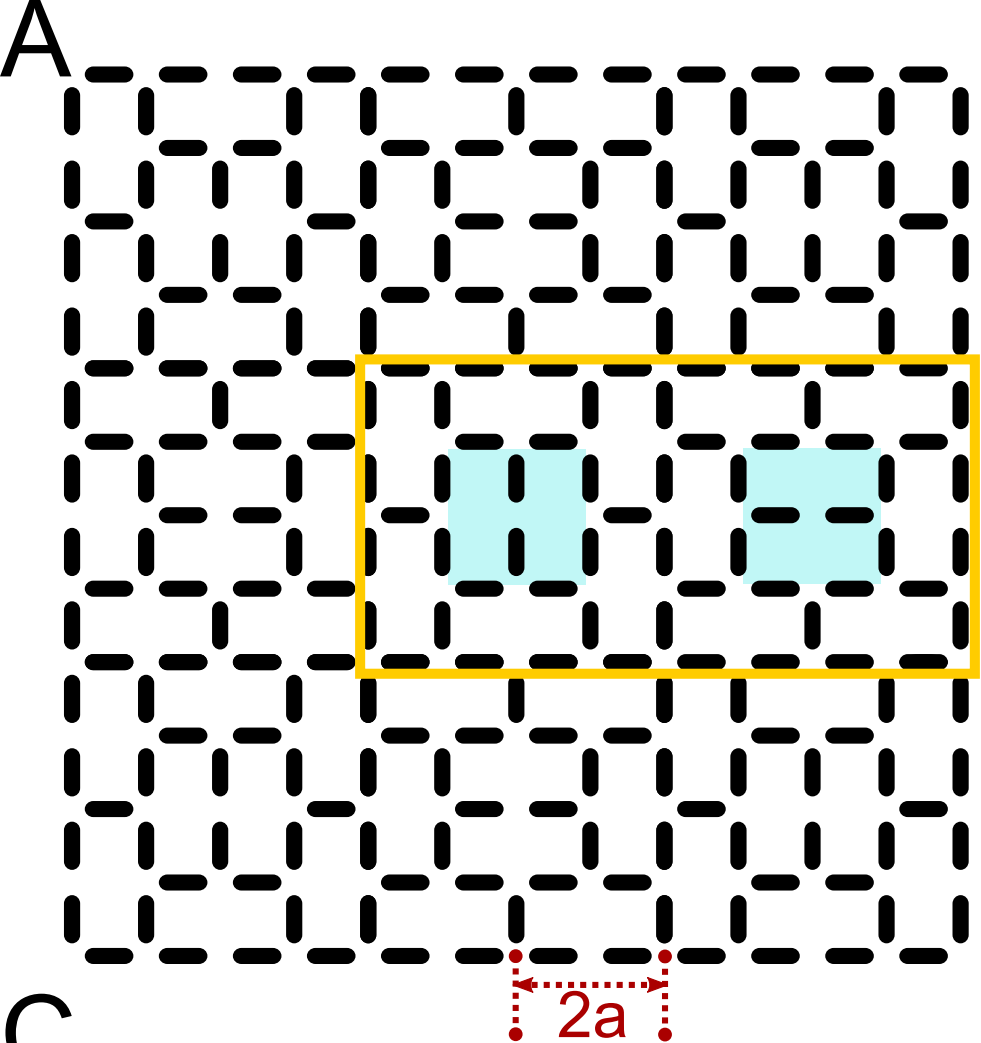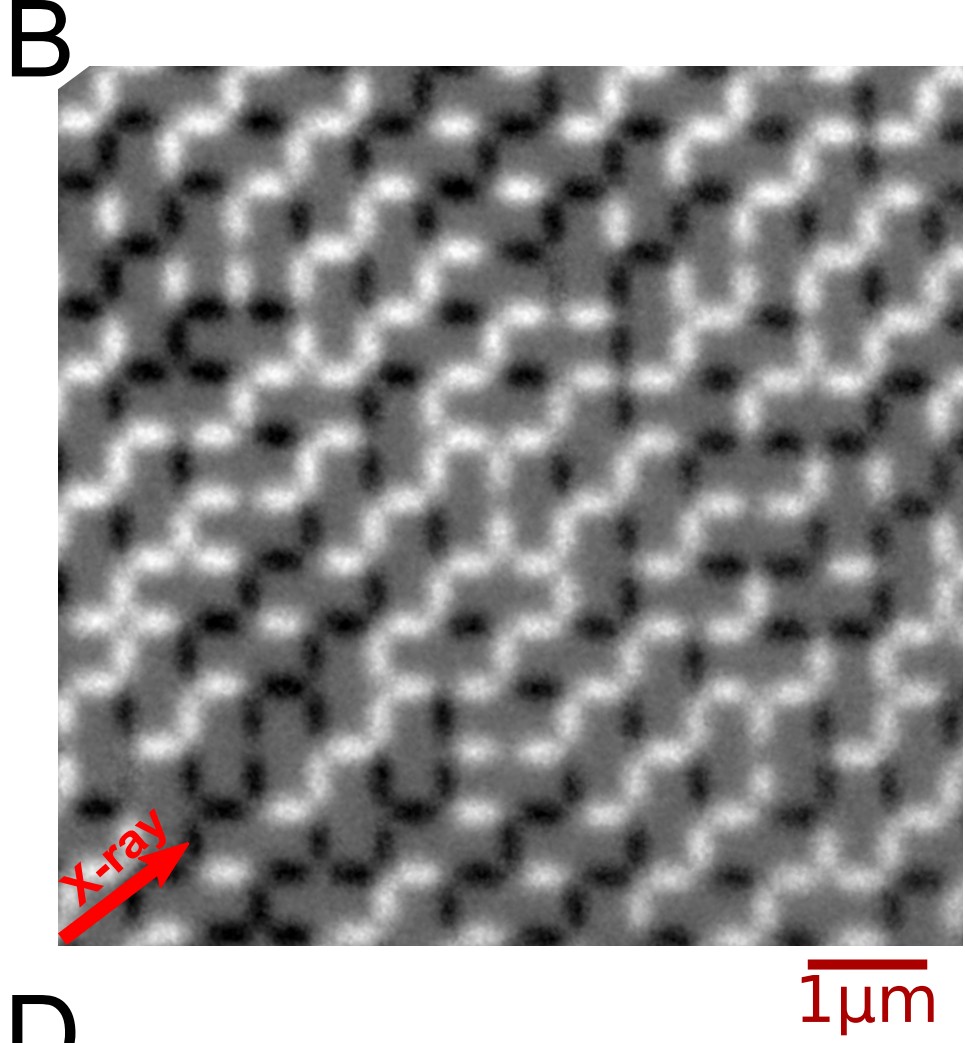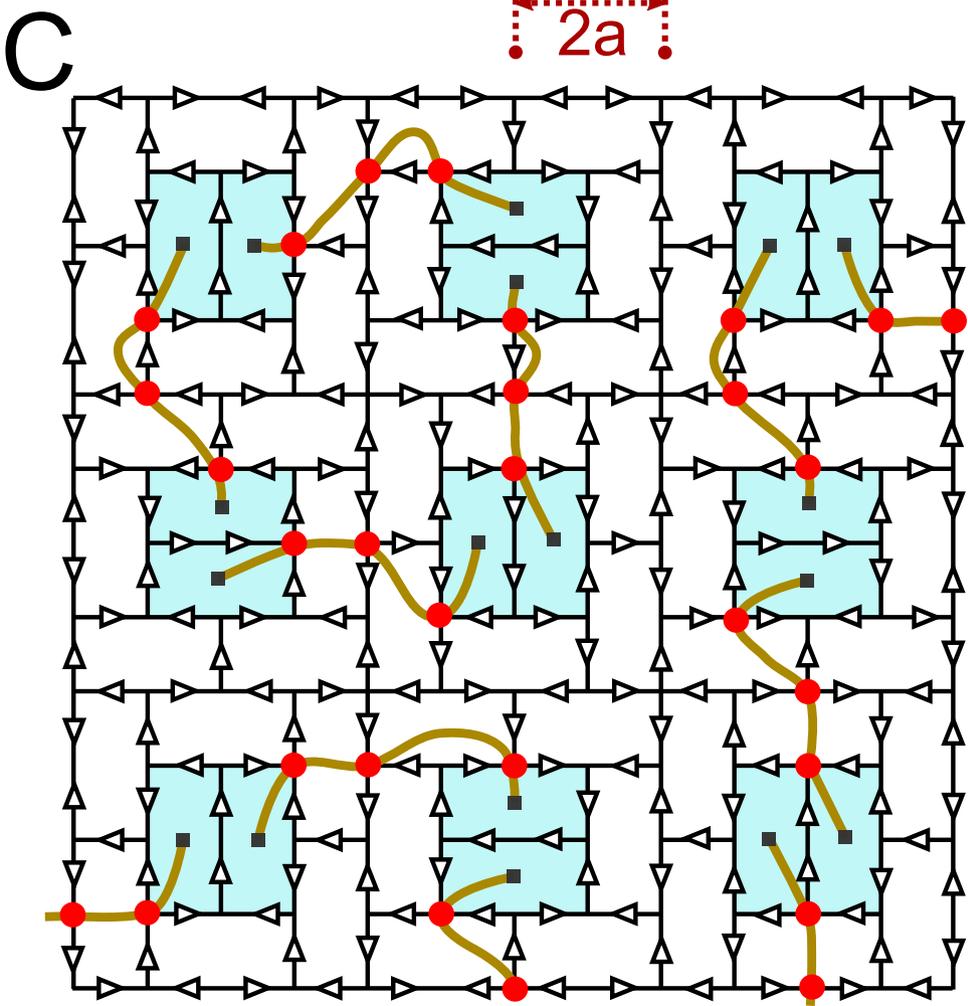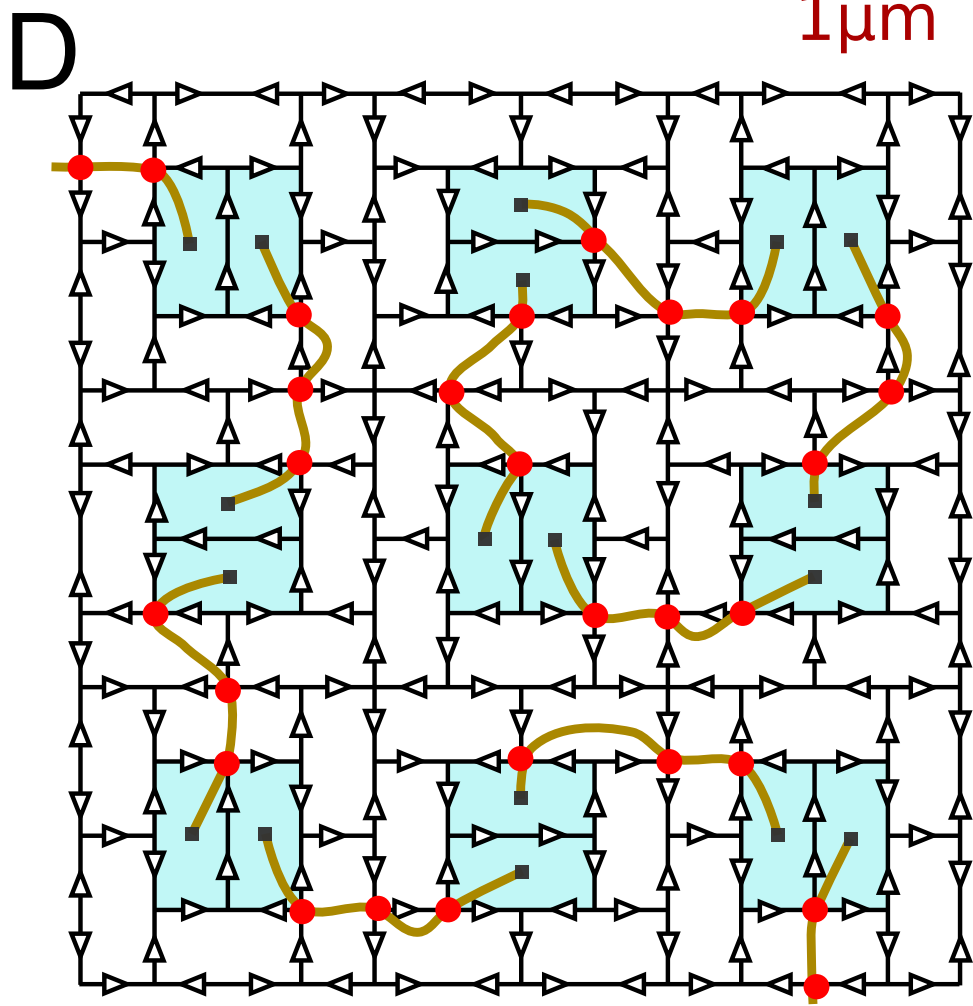

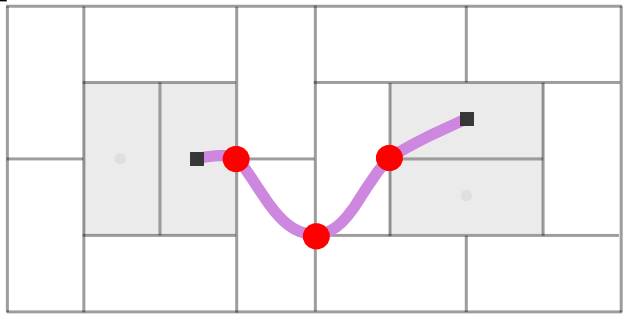 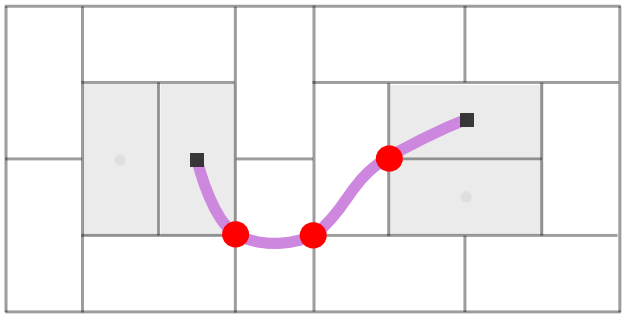 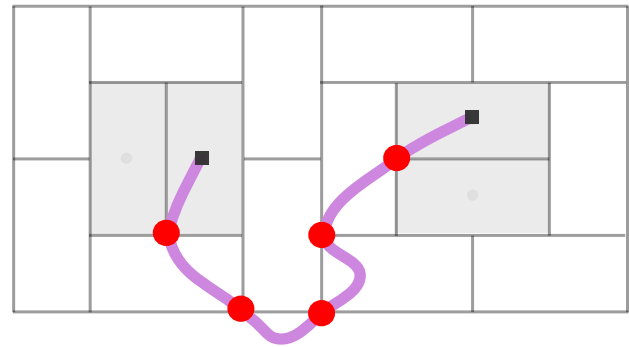
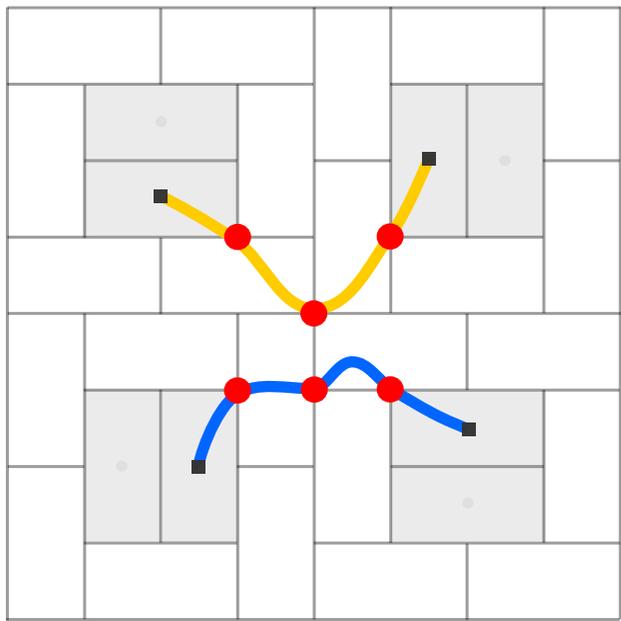 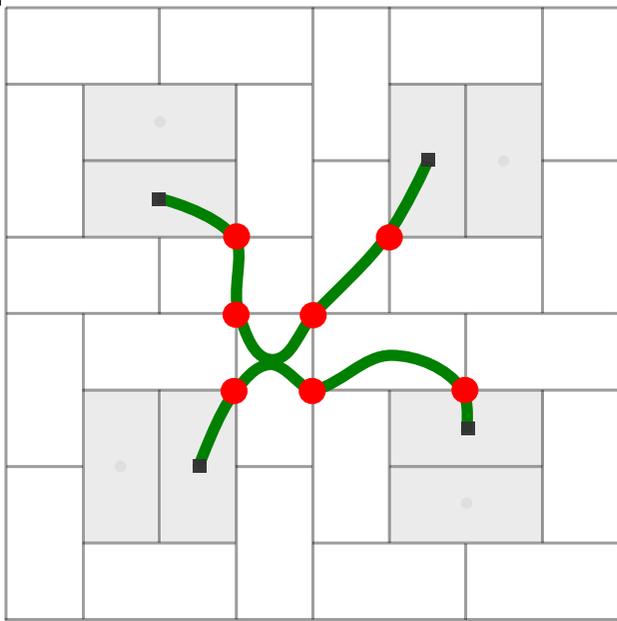 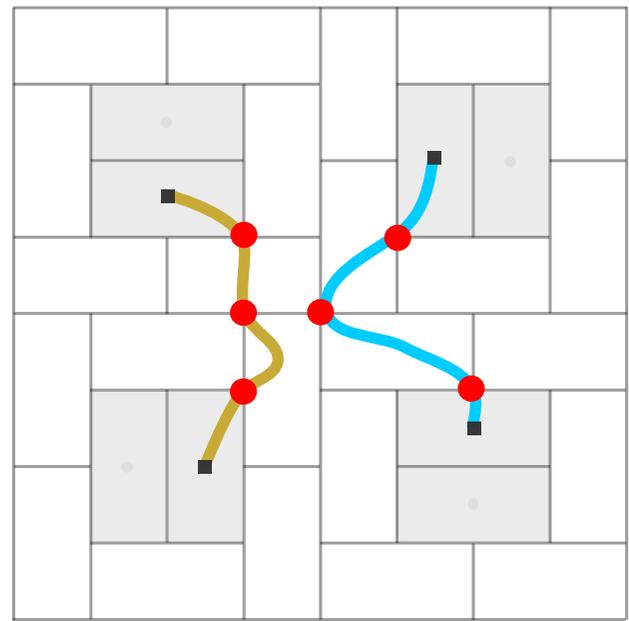

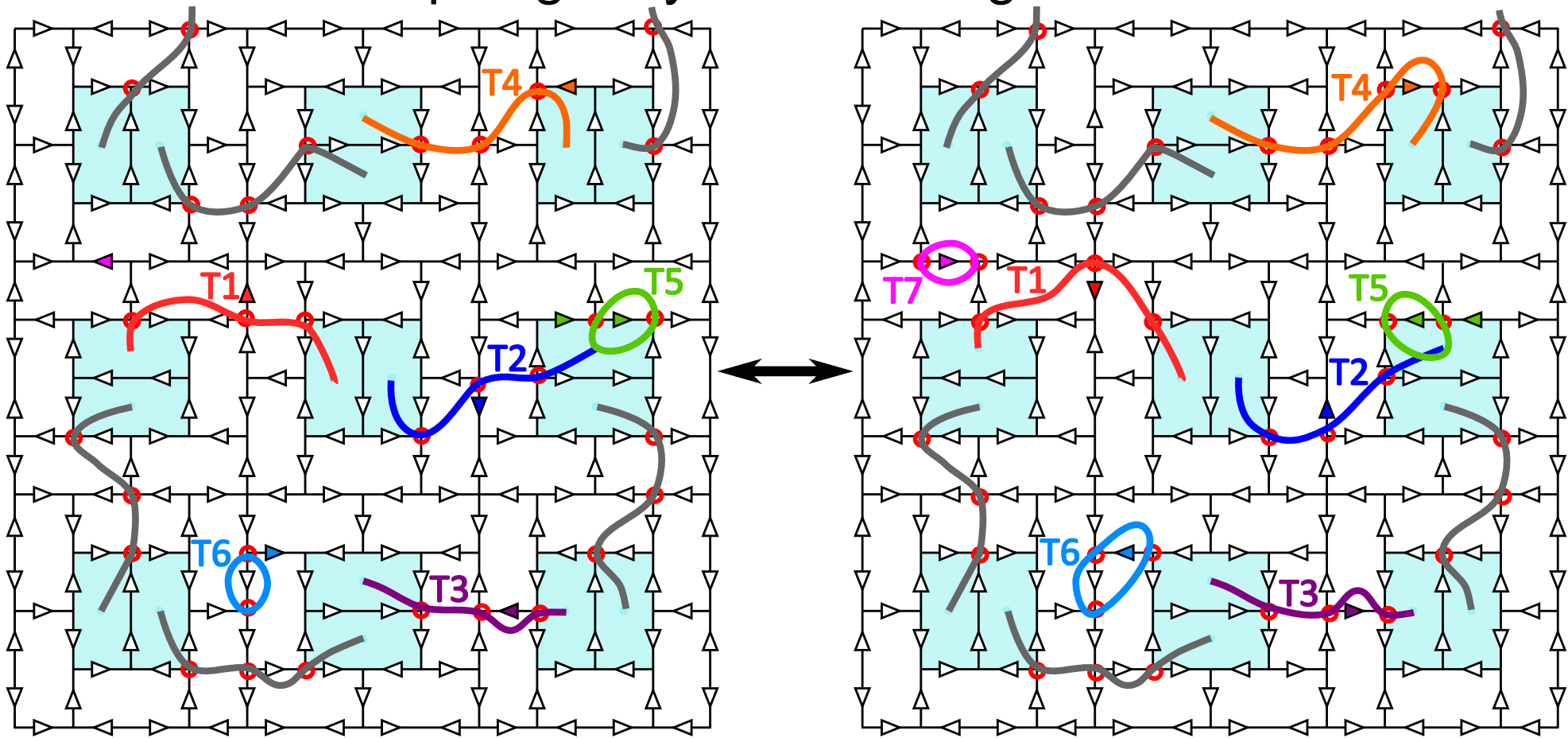

## Topologically Trivial String Motions

| Topologically trivial motions | | |
|---|---|---|
| Wiggle | | Energy increase (T1) |
| | | Energy decrease (T1) |
| | | No energy change (T2,T3) |
| Grow (T4) | | |
| Shrink (T4) | | |
| Loop | | Wiggle (T5) |
| | | Grow (T6) |
| | | Shrink (T6) |
| | | Creation (T7) |
| | | Annihilation (T7) |

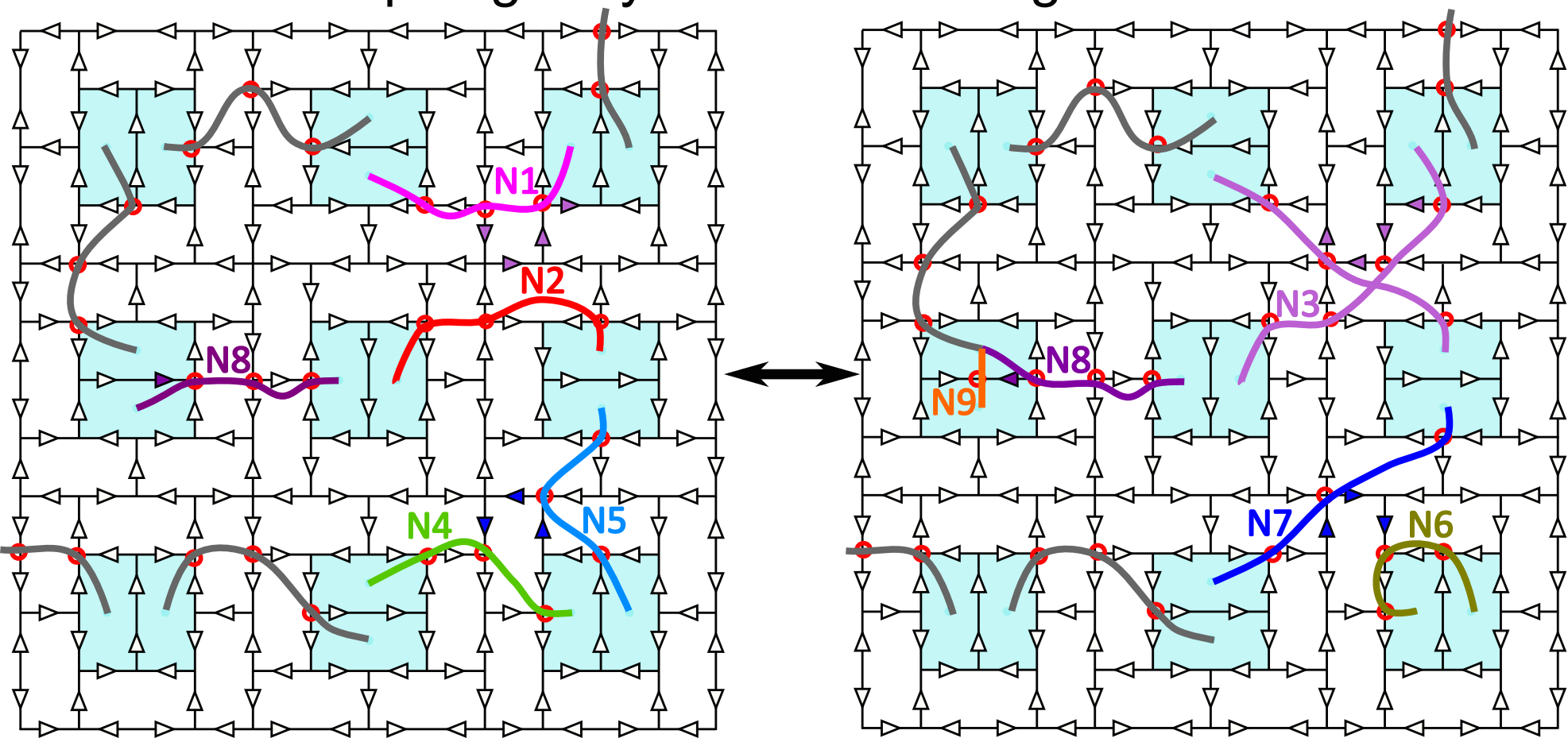

## Topologically Nontrivial String Motions

| Topologically nontrivial motions | |
|---|---|
| Merge (N1,N2→N3) | |
| Split (N3→N1,N2) | |
| Reconnection (N4,N5↔N6,N7) | |
| Adjacent reconnection (N8) | |
| Strings on adjacent interior plaquette | Creation (N9) |
| | Annihilation (N9) |

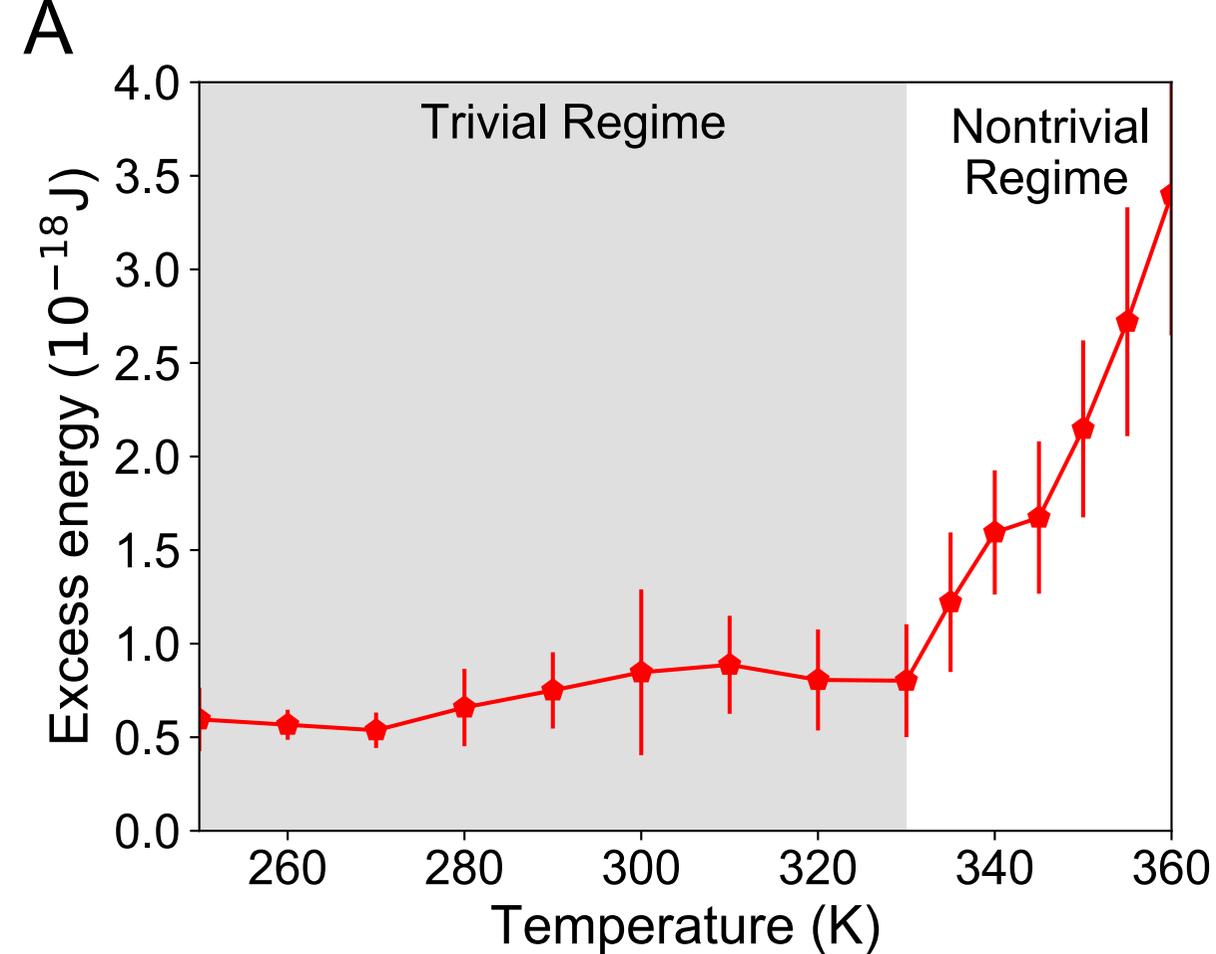
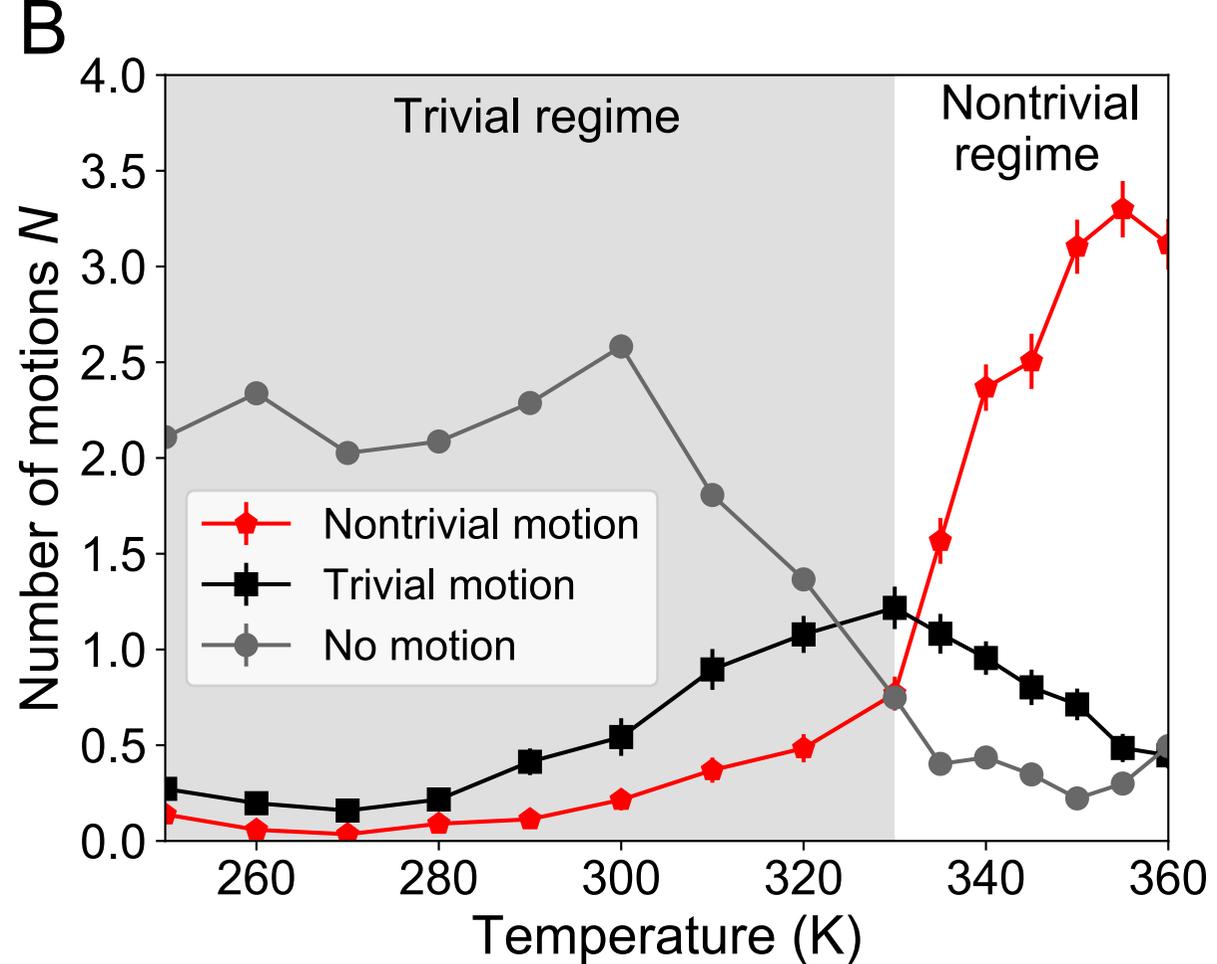
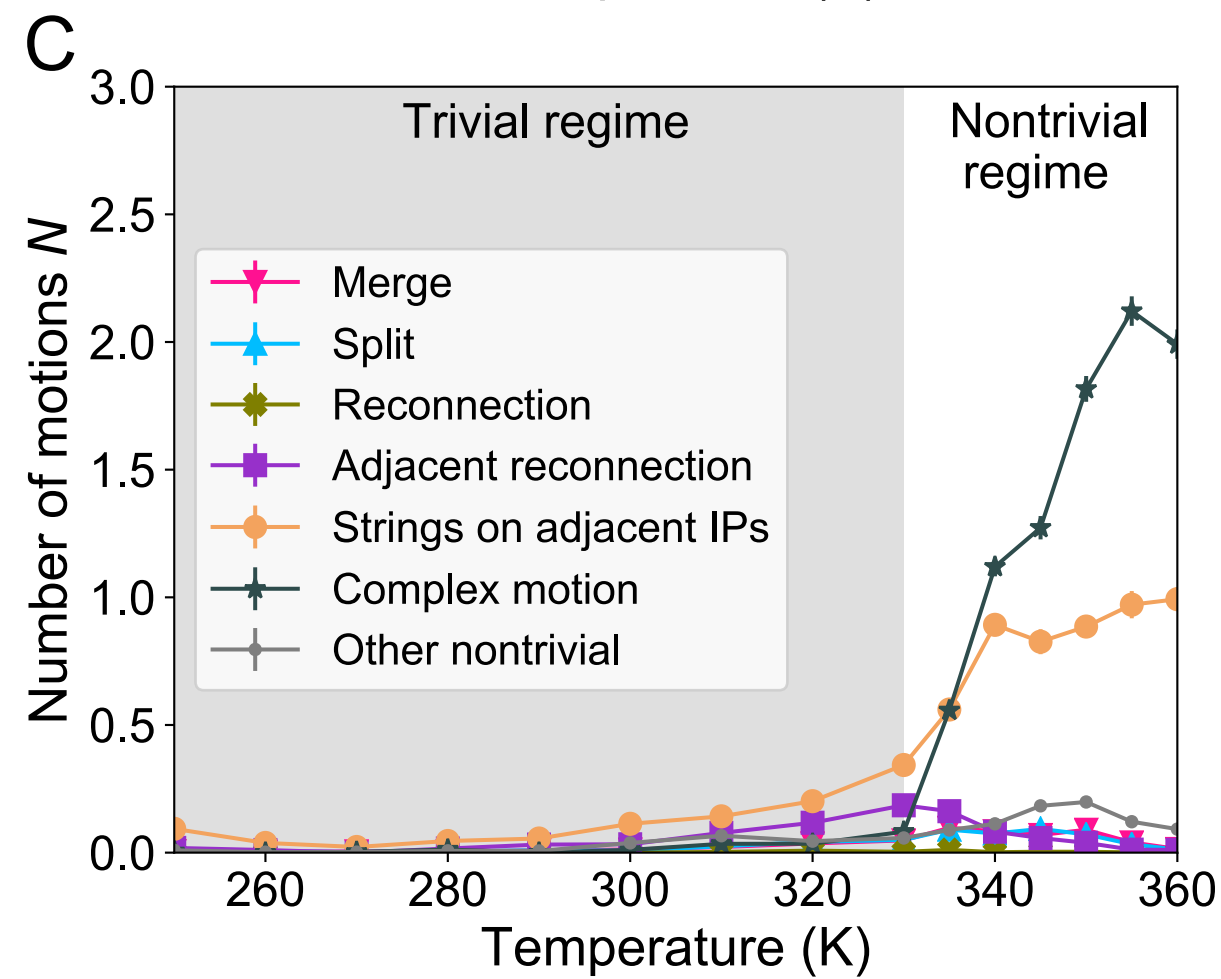
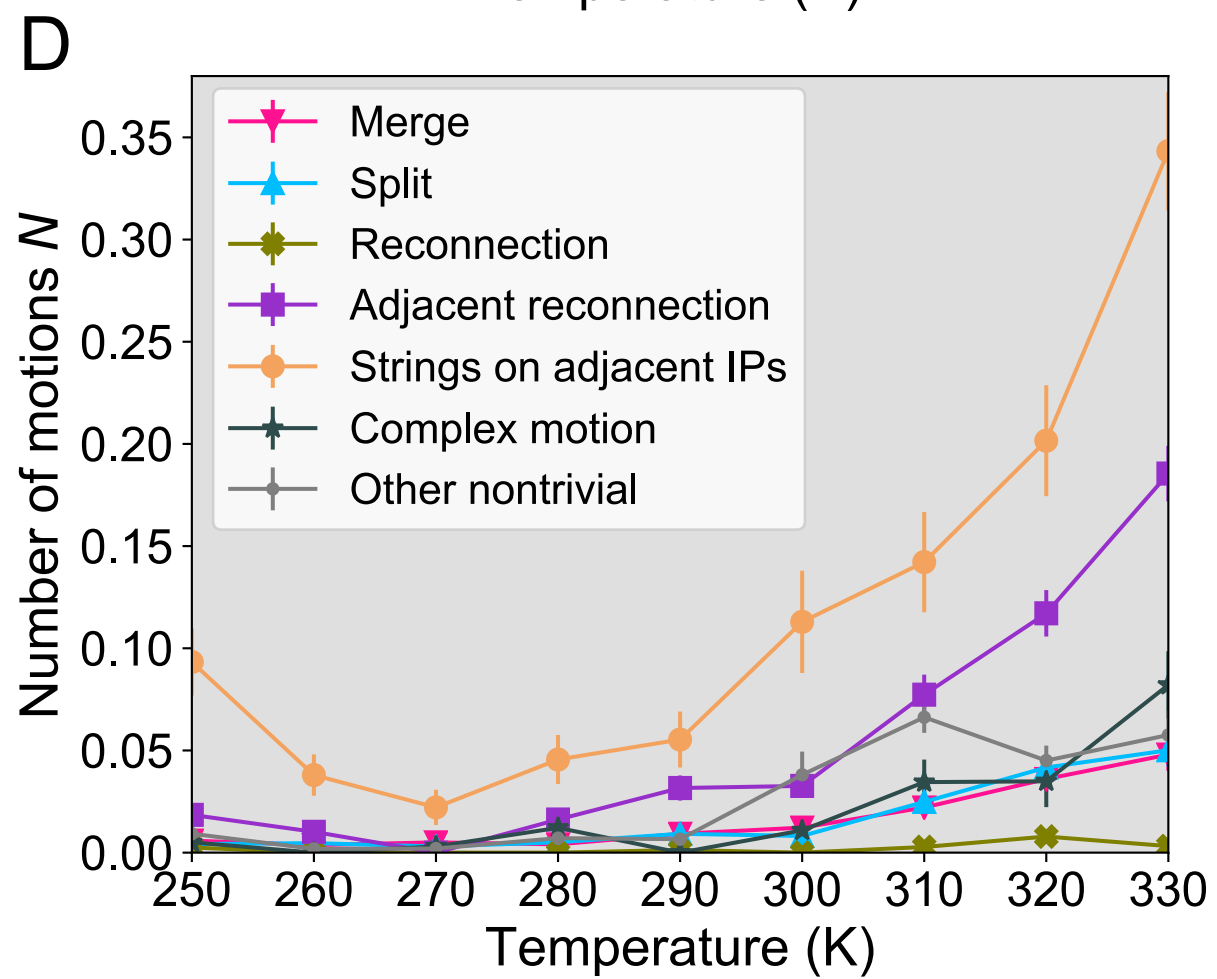

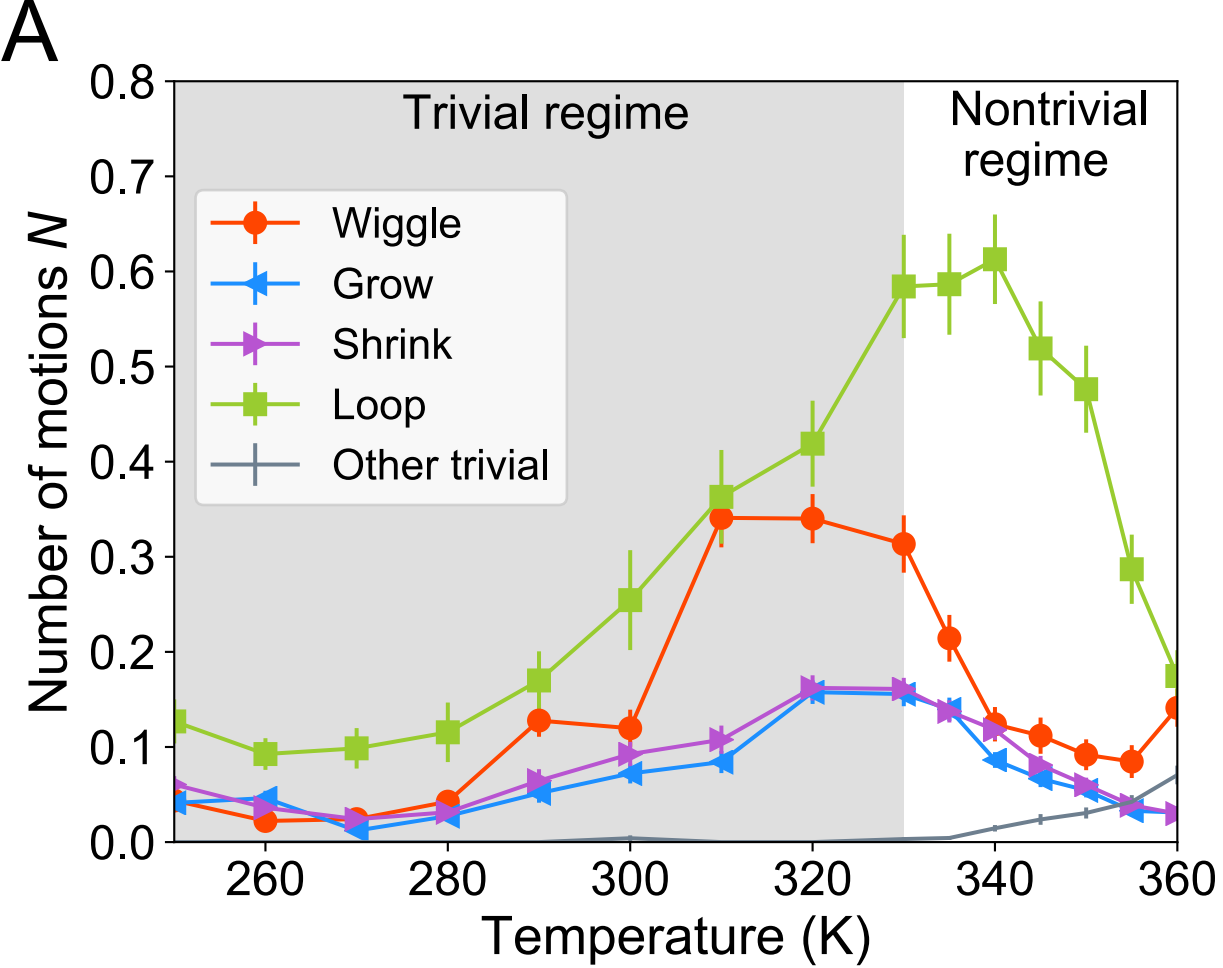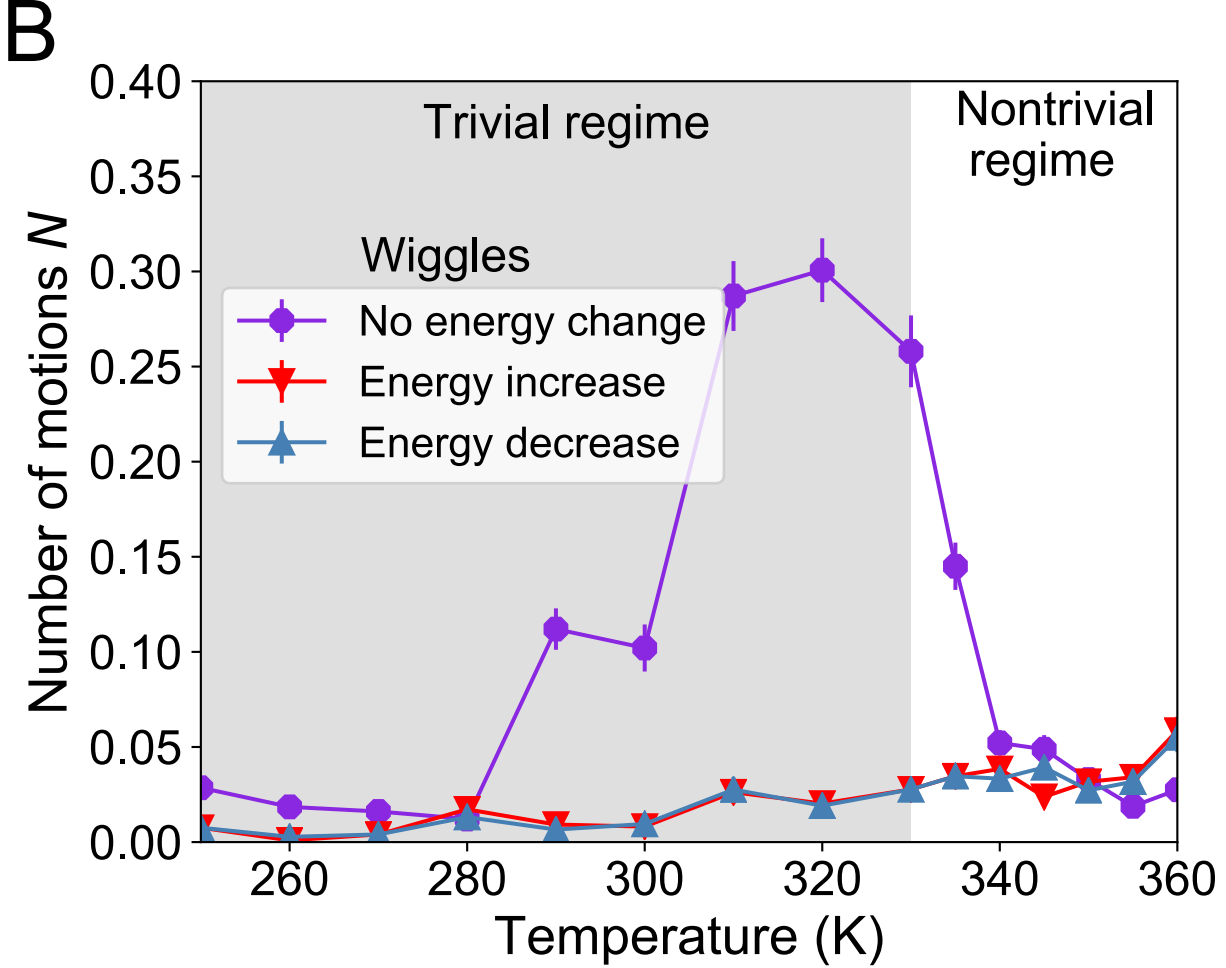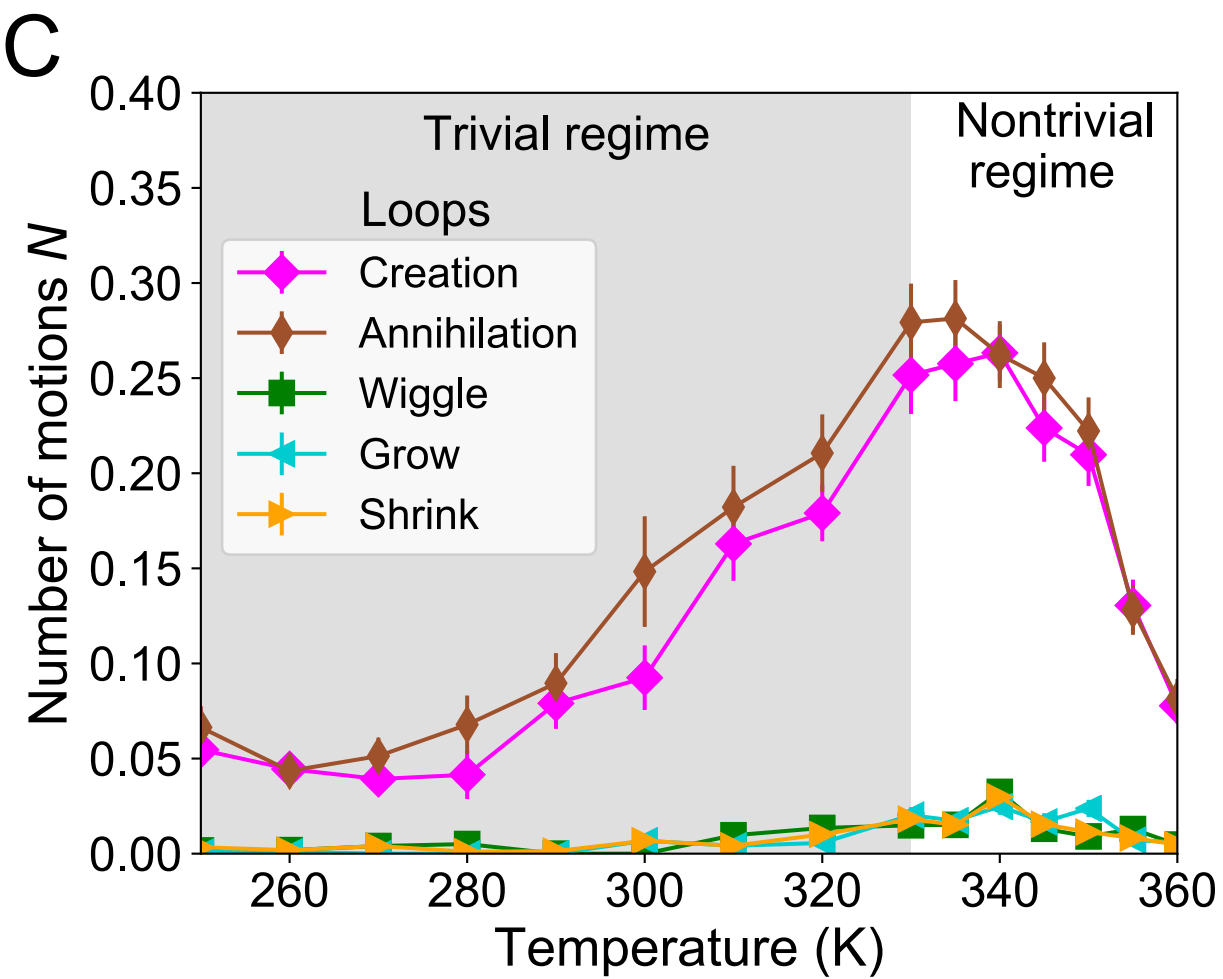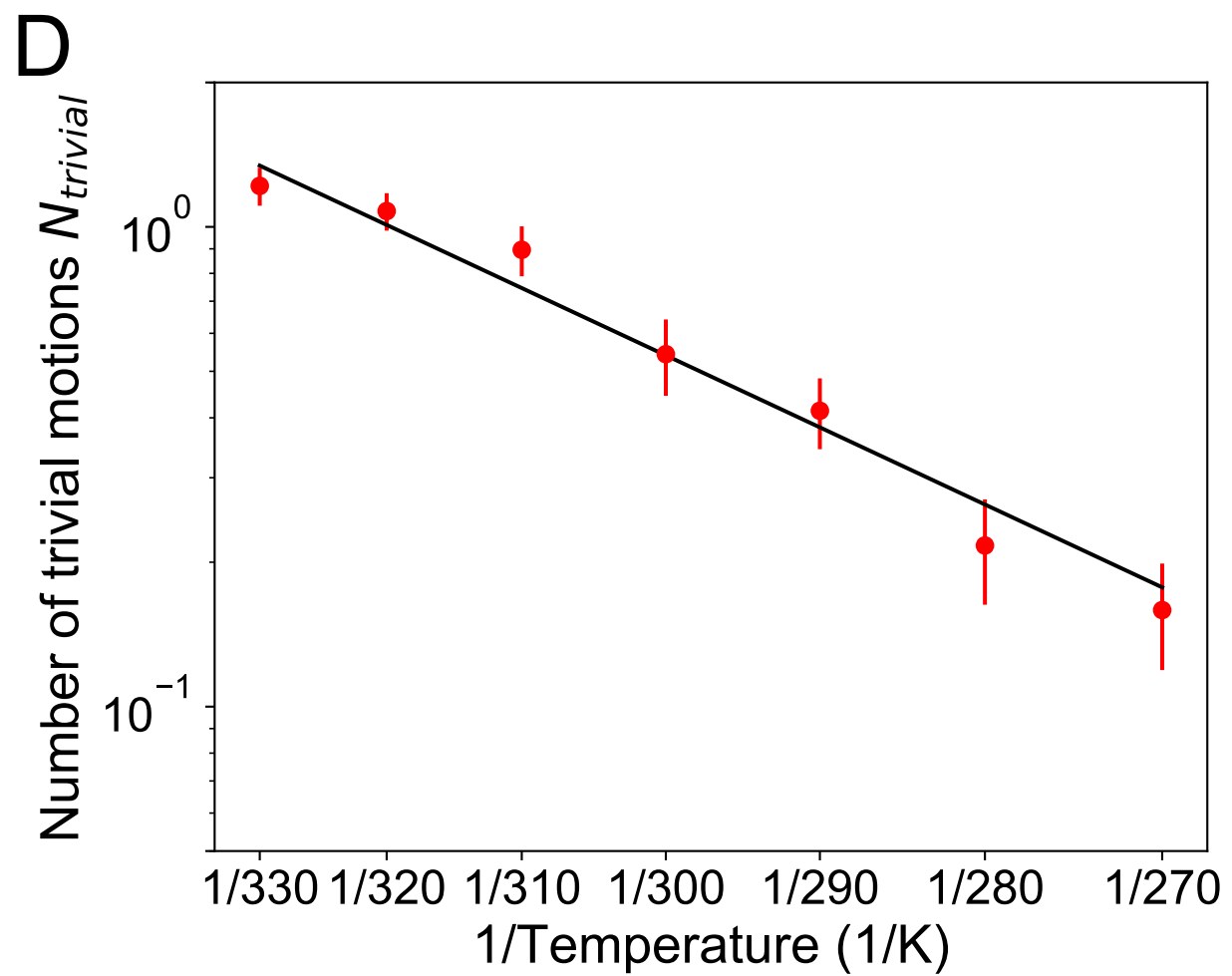

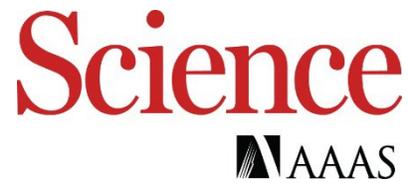

# Supplementary Materials for

**Topological Kinetic Crossover in a Nanomagnet Array**

Xiaoyu Zhang[1], Grant Fitez[1], Shayaan Subzwari[1], Nicholas S. Bingham[1], Ioan-Augustin Chioar[1], Hilal Saglam[1], Justin Ramberger[3], Chris Leighton[3], Cristiano Nisoli*[2], & Peter Schiffer*[1]

Correspondence to: peter.schiffer@yale.edu and cristiano@lanl.gov

**This PDF file includes:**

Materials and Methods
Supplementary Text
Figs. S1 to S6
Tables S1 to S2

**Materials and Methods**

S1. Sample preparation

The specifics of sample fabrication and characterization have been described in detailed in a previous paper [$14$], where we also showed that the string population and length distributions are thermally activated at the highest experimentally accessible temperatures. We first used electron-beam lithography to write patterns on Si/SiO$_x$ substrates with spin-coated bilayer resists. The patterns were made of islands with designed lateral dimensions of 470 nm × 170 nm, and lattice spacings of 600 nm, 700 nm, and 800 nm. A layer of ~ 2.5 nm thick permalloy (Ni$_{80}$Fe$_{20}$) was then deposited into the patterns via ultrahigh vacuum electron beam evaporation, followed by aluminum capping layers of 2 nm thick to prevent oxidation of the underlying permalloy.

S2. PEEM-XMCD imaging

We measured the moment configuration in our samples using x-ray magnetic circular dichroism photoemission electron microscopy (XMCD-PEEM), which yields real-space images of the island magnetic moment. The experiments were conducted at the PEEM-3 endstation at beamline 11.0.1.1 of the Advanced Light Source, Lawrence Berkeley National Lab. All data are derived from the same XMCD-PEEM images taken on the same samples that were analyzed in reference [14].

We first cooled the samples down to 250 K where the island moments were thermally stable. We then heated the sample from 250 K to 360 K in 5-10 K steps and took 100 PEEM images at the Fe L$_3$ absorption edge at each temperature point. The 100 PEEM images consisted of ten exposures with a left-circularly polarized X-ray beam followed by ten exposures with a right-circularly polarized beam, repeated five times. The data acquisition sequence at each temperature is therefore five repeats of "LLLLLLLLLL [5 second switch] RRRRRRRRRR [5 second switch]", where L and R correspond to images with different polarization of the X-rays. Each image takes 0.5 seconds, and then there is a delay of 0.5 seconds between images (assuming polarization is not switched), so the time period of the imaging is 1 second. The total acquisition time at each temperature was about 150 seconds including computer read-out between exposures and switching x-ray polarization.

The PEEM image field of view was set at 15 x 15 to 18 x 18 μm$^2$ and there were about 500 islands within each image. Upon increasing temperature, thermal excitations caused a fraction of the moments to visibly flip in orientation, with the rate of flipping increasing with increasing temperature. We used MATLAB and Python code to extract the moment direction for each island and emergent strings configuration from every PEEM image. By comparing the string configurations between successive PEEM images, we could therefore observe the thermal kinetics of this system.

S3. String representation

The emergent string representation has previously been described in reference [$14$] and is summarized here. Fig. S1. (A) gives examples of how the strings transect the type II excited states of 2-island and 3-island vertices and the type II and III excited states of 4-island vertices (the different coordination vertices are typically labelled as $z = 2,3,4$ and given as subscripts in the figure). On the vertex level, if a pair of islands on a vertex are aligned as the energetically favorable configuration (moments M1 and M3 on the $z = 3$ vertex for example), there is no string segment between them. Otherwise, there is a segment of string that separates a pair of moments that are not in their energetically minimum configuration, for instance, between M1 and M2 or between M2 and M3.

Beyond the vertex level, the SFI structure is made of the interior (shaded in pink in Fig. S1. (B)) and peripheral plaquettes. The interior plaquettes each are bounded by a $z = 2$ vertex and five $z = 3$ vertices, while the peripheral plaquettes only have $z = 3$ and $z = 4$ vertices around them. As explained in reference [*14*], because of the presence of a $z = 2$ vertex, the interior plaquettes necessarily have an odd number of excited vertices connecting two of them, whereas the peripheral plaquettes always have an even number of excited vertices, and they can have zero excited vertices. Thus, after forming the full set of strings by connecting the segments of the strings from individual vertices that are shown in Fig. S1. (A), excluding strings affected by sample boundaries, all resulting strings must either anchor on the interior plaquettes or form closed loops within peripheral plaquettes, as illustrated in examples shown in the Fig. S1. (B).

S4. Algorithm for detecting string motions
As noted above, all strings are either loops or have two ends located in interior plaquettes (IPs). The string motion analysis program starts with labeling each string with two sets: one set is the center of each of the IPs at the string's ends, and the other set is the excited vertices that the string connects. The string motions are only measured between consecutive LL or RR pairs of images, excluding a small fraction of bad images and images at the beginning or end of the set of 10 L or 10 R images.

Beginning with the second XMCD-PEEM image in a group of images with the same x-ray polarization, we compared the strings with the previous XMCD-PEEM image, and classified the string motions in the following scheme:

1. No change: Neither the IPs in the set nor the excited vertex sets change.
2. Trivial motion: The IPs in the set do not change, while the excited vertices do change.
    a. Wiggle: The number of excited vertices does not change.
        i. Energy increase: The total energy of the excitations is higher.
        ii. Energy decrease: The total energy of the excitations is lower.
        iii. No energy change: The total energy of the excitations does not change.
    b. Grow: The number of excited vertices increases.
    c. Shrink: The number of excited vertices decreases.
    d. Loops: There are no IPs in the set. (Even if a loop touches an IP, it is still counted as a loop.)
        i. Loop created: The loop is absent in the previous image.
        ii. Loop annihilated: There is a loop in the previous image, but it is absent in the current image.
        iii. Loop wiggle: The number of excitations does not change.
        iv. Loop grow: The number of excitations increases.
        v. Loop shrink: The number of excitations decreases.
    e. Ambiguous trivial motion: When two or more strings share the same set of IPs, all we know is the number of strings and it is hard to further distinguish between them. For example, if two strings share the same anchor points in one image, after some islands flip, one undergoes a trivial motion and the other undergoes a nontrivial motion. Since we do not know which string is which, these motions are ambiguous.

3. Non-trivial motion: The IPs in the set change.
    a. Merge: The sets of IPs from two strings join.
    b. Split: The set of IPs of one merged string separates.
    c. Reconnection: The IPs from two strings redistribute into two new strings.
        i. Adjacent reconnection: One IP on a string switch to the adjacent IP that shares a $z = 2$ vertex with the original IP.
        ii. Other reconnections.
    d. New strings created/annihilated:
        i. Strings on adjacent IP: a string that connects two IPs sharing a $z = 2$ vertex.
        ii. Other new strings created/annihilated.
    e. Ambiguous nontrivial motion.
    f. Complex motion: A motion involving multiple string motions as shown in Fig. S2.
    g. Incomplete string nontrivial motion: The IP set changes but the string motion does not satisfy any one of the previous requirements because of boundaries or missing islands in the image.

The incomplete strings, that only attach to 1 or 0 IP and do not form a loop, are discarded for our analysis. The incomplete strings are either on the edge of frames where the IP(s) is(are) outside the field of view or are discontinuous strings because of unrecognizable islands in the middle of strings. In either one of these cases, "new" string creation or annihilation may be recognized through island flips. However, these "new" strings are produced because of the limitation of image size or analysis program; they are not real "new" strings, therefore, should not be considered.

**Supplementary Text**

S5. Additional XMCD-PEEM data

We performed XMCD-PEEM measurements on $a$ = 600 nm, 700 nm and 800 nm SFI. The data for $a$ = 600 nm SFI are presented in the main text. The $a$ = 700 nm and 800 nm SFI samples have consistent results as shown in Fig. S3. and S4. respectively. And the total length of strings and loops are shown in Fig. S5.

The crossover temperatures are $T_X$ ~ 330 K for $a$ = 700 nm sample and ~ 320 K for $a$ = 800 nm sample. For larger lattice constant samples, the nanoislands interact more weakly, consistent with a lower crossover temperature.

The data for the number of trivial processes versus the reciprocal temperature for $a$ = 700 nm and 800 nm both show apparent thermally activated behavior, consistent with the results in the main text for $a$ = 600 nm. We performed a nonlinear fit to the data for $N_{trivial}$ as a function of $1/T$, to the form given in the main text. The fit parameters again depend strongly on the choice of fitting range, with $t_o$ ~ $10^{-7}$ - $10^{-8}$ s and $E_S$ ~ 5000 K for $a$ = 700 nm and $t_o$ ~ $10^{-5}$ s and $E_S$ ~ 3000 K for the $a$ = 800 nm. The non-monotonic dependence on the fit parameters on lattice spacing is consistent with behavior seen in the simple island fluctuation rate, published previously [*14*].

S6. Micromagnetic Energy Calculations

Permalloy films have in-plane magnetization and weak magnetocrystalline anisotropy, and thus the island shape dictates the energy barrier to flipping each island moment. To estimate the energy barrier of single island moment flipping, we calculated the zero-temperature magnetostatic energy of configurations with moments along the long axis and the short axis. We define the flipping barrier as the energy difference between two configurations divided the Boltzmann constant $k_B$. We have calculated the magnetostatic energy and energy barrier of three SFI lattices using the micromagnetic simulation program MuMax3 [28]. While doing the simulations, the saturation magnetization is set to be $8.6 \times 10^5$ A/m, the exchange constant is $1.3 \times 10^{-11}$ J/m, and the magnetocrystalline anisotropy is negligible for permalloy, which is consistent for previous work [14]. Table S1. Gives the nanoisland dimensions measured from SEM and the energy barrier of three SFI lattices.

SFI contains three types of vertices: $z = 2$, 3, and 4 vertices and each has two to four possible moment configurations as shown in Fig. S6. We have calculated the energy of all the moment configurations for $a = 600$ nm, 700 nm, and 800 nm SFI arrays using the same program. Table S2. gives the ground state energies for all three types of vertices and the excess energy above ground states for all the excited states. The excess energy for a given type of vertex is defined as the energy above the possible ground state configurations (i.e., the excess energy of a $z = 3$ type II vertex is $\Delta E_{II}^3 = E(II_3) - E(I_3)$).

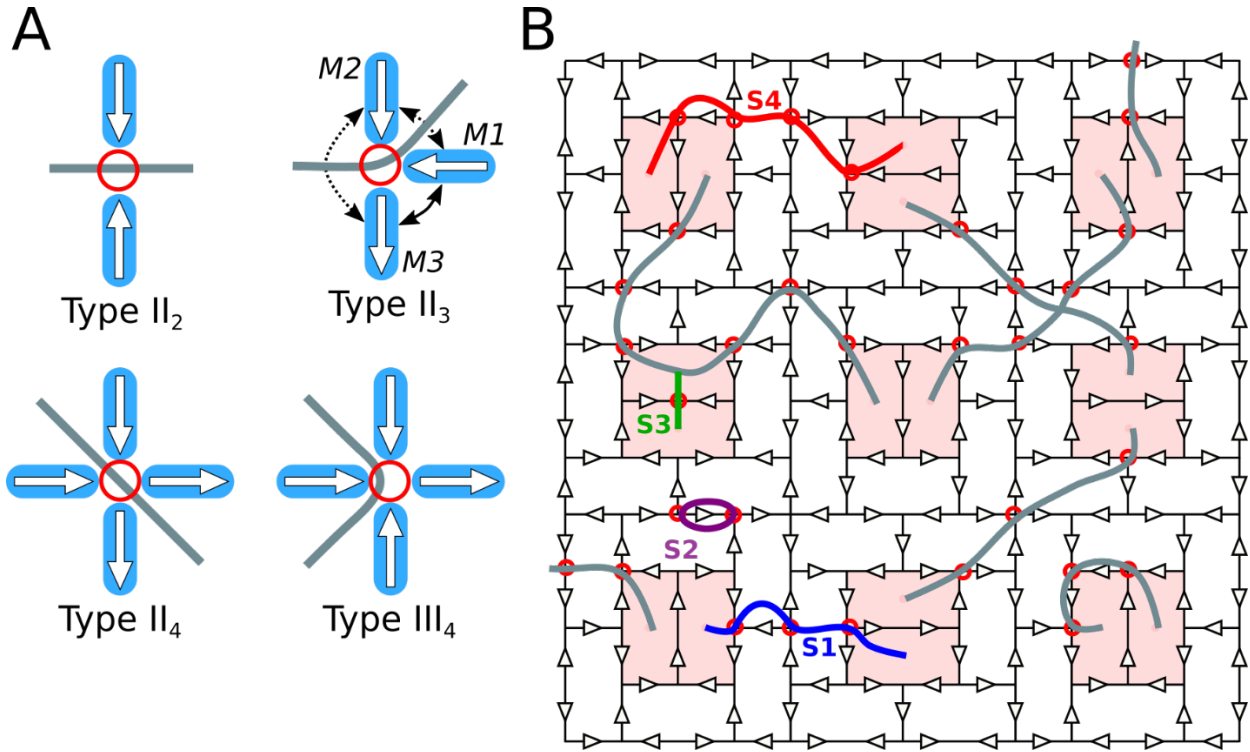

**Fig. S1. String representation on vertices and SFI structure.** (A) Segments of strings on selected excited vertices. The solid line with arrows on type $II_3$ indicates that island moments M1-M3 are aligned energetically favorably, while the dashed lines indicate the energetically unfavorable alignment between M1-M2 and M2-M3. (B) Strings on SFI lattices with moment configuration indicated by the open arrow heads. All strings are either loops (the purple loop S2) or have two ends located in an interior plaquette. S1 connects two non-adjacent interior plaquettes with length of the minimum of three excited vertices. S3 connects two adjacent interior plaquettes. S4 connects two non-adjacent interior plaquettes with length of four excited vertices.

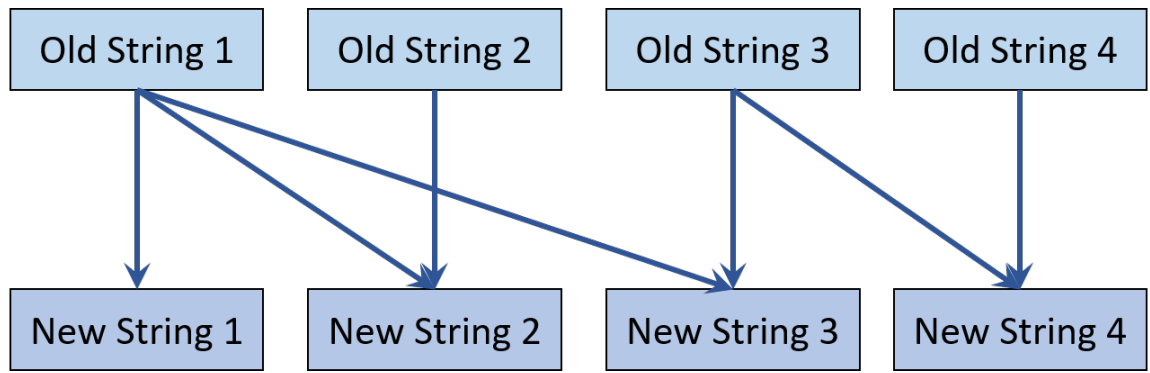

**Fig. S2.** An example of complex motion that consist of two splits and 3 merges.

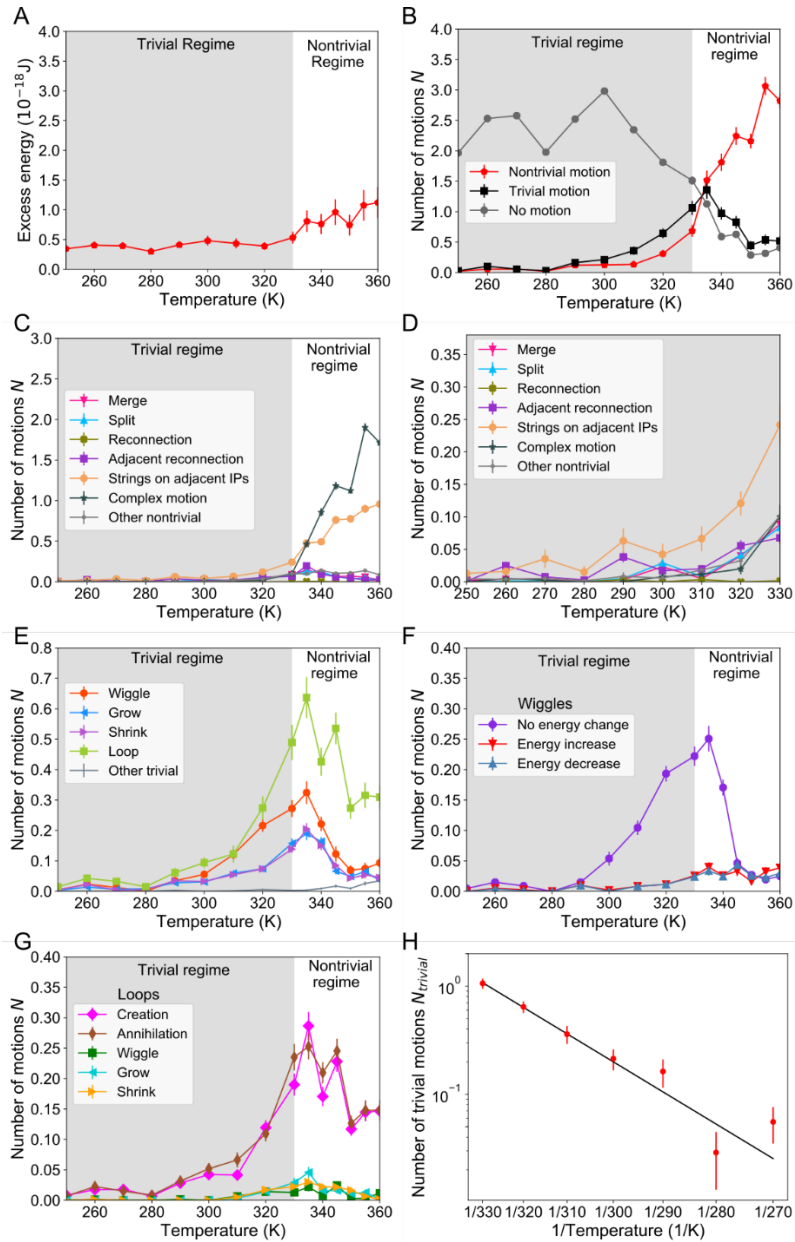

**Fig. S3. Temperature dependent string properties for $a$ = 700 nm SFI.** (A) The excess energy per unit cell versus temperature. (B) The temperature dependence of the number of nontrivial (red) and trivial (black) string motions per unit cell. The number of no motion strings is given in dark gray. (C) Temperature dependence of all types of nontrivial motions and (D) an expanded plot of the low temperature regime. (E) All types of trivial motions. (F) String wiggle subtypes of trivial motions. (G) Loop subtypes of trivial motions. (H) The number of total trivial motions as a function of inverse temperature with fit. The error bars are the standard errors from all XMCD-PEEM images taken at given temperature.

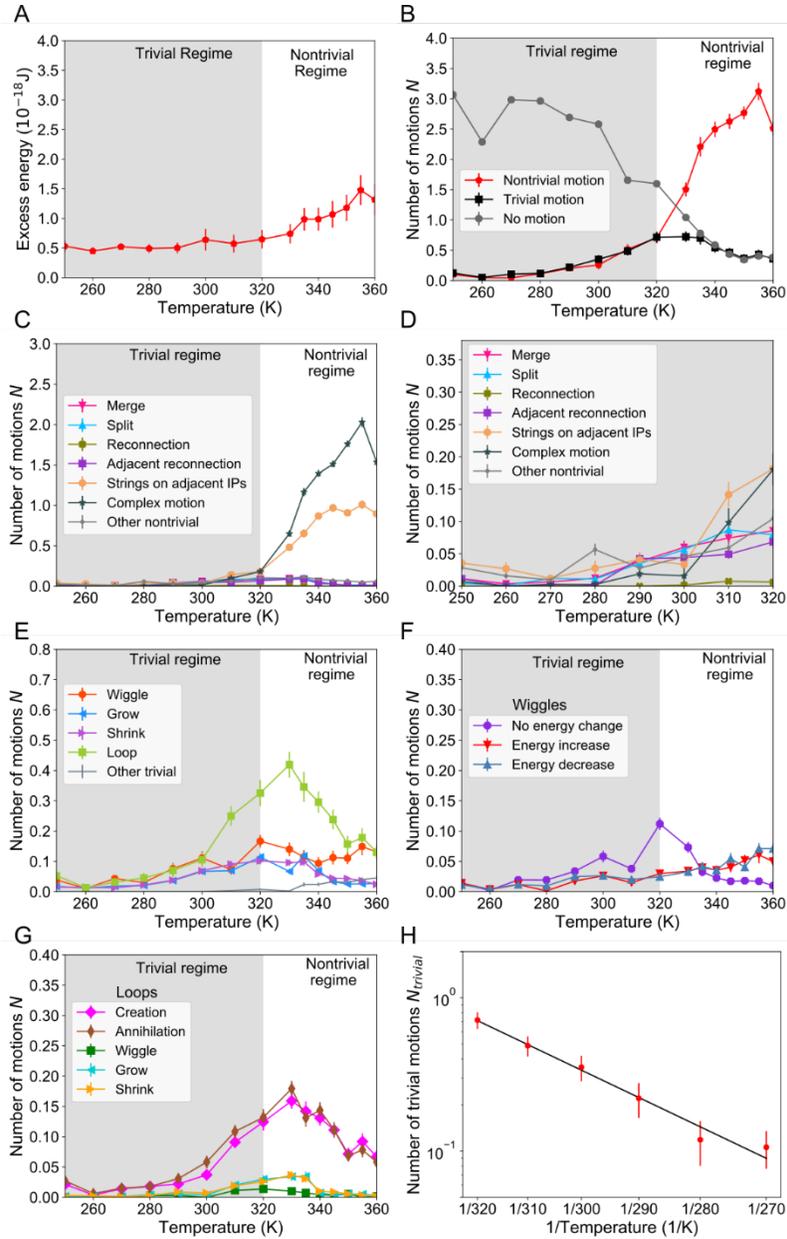

**Fig. S4. Temperature dependent string properties for $a$ = 800 nm SFI.** (A) The excess energy per unit cell versus temperature. (B) The temperature dependence of the number of nontrivial (red) and trivial (black) string motions per unit cell. The number of no motion strings is given in dark gray. (C) Temperature dependence of all types of nontrivial motions and (D) an expanded plot of the low temperature regime. (E) All types of trivial motions. (F) String wiggle subtypes of trivial motions. (G) Loop subtypes of trivial motions. (H) The number of total trivial motions as a function of inverse temperature with fit. The error bars are the standard errors from all XMCD-PEEM images taken at given temperature.

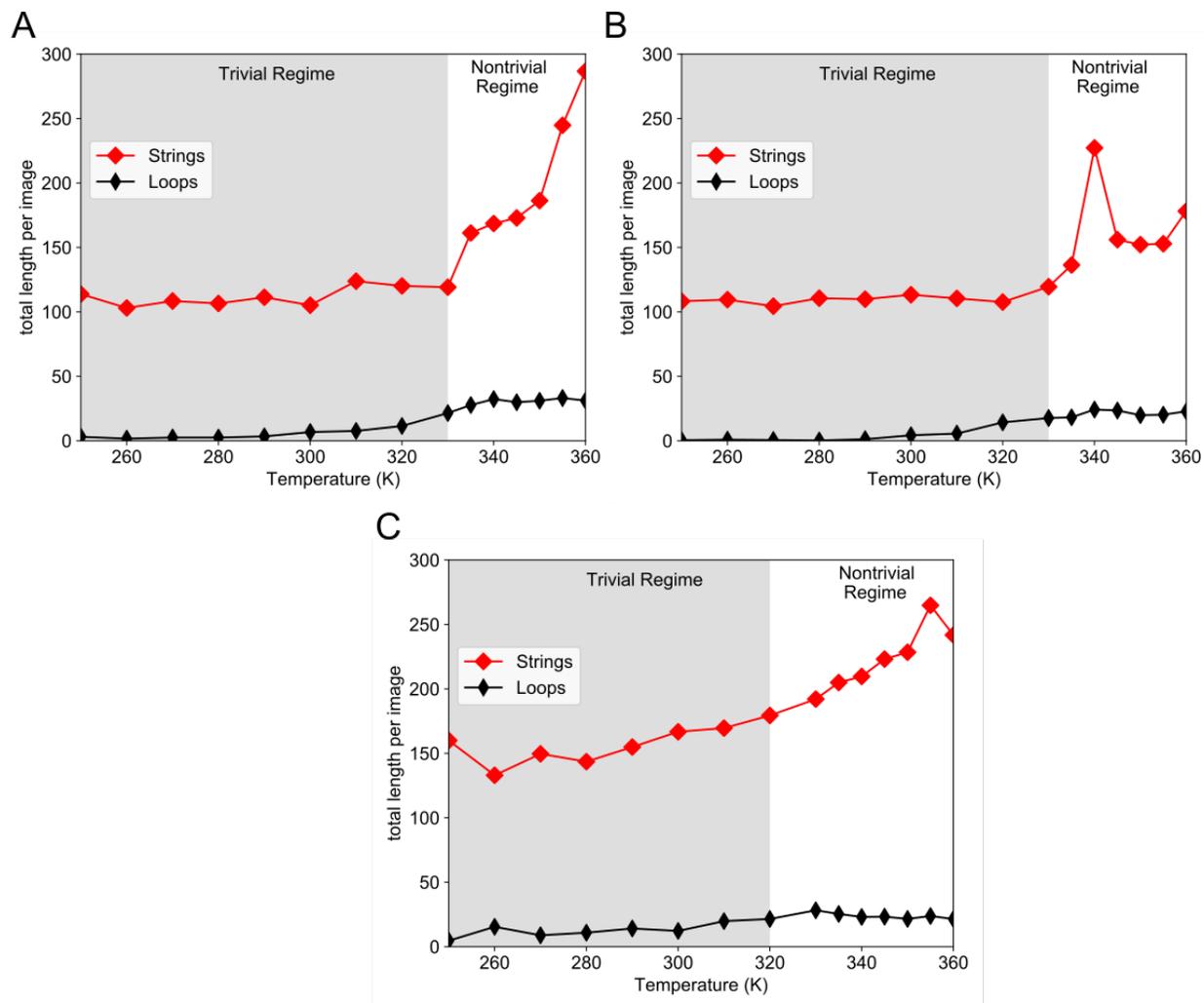

**Fig. S5.** The total length of loops and non-loop strings as a function of temperature for $a =$ (A) 600 nm, (B) 700 nm, and (C) 800 nm SFI.

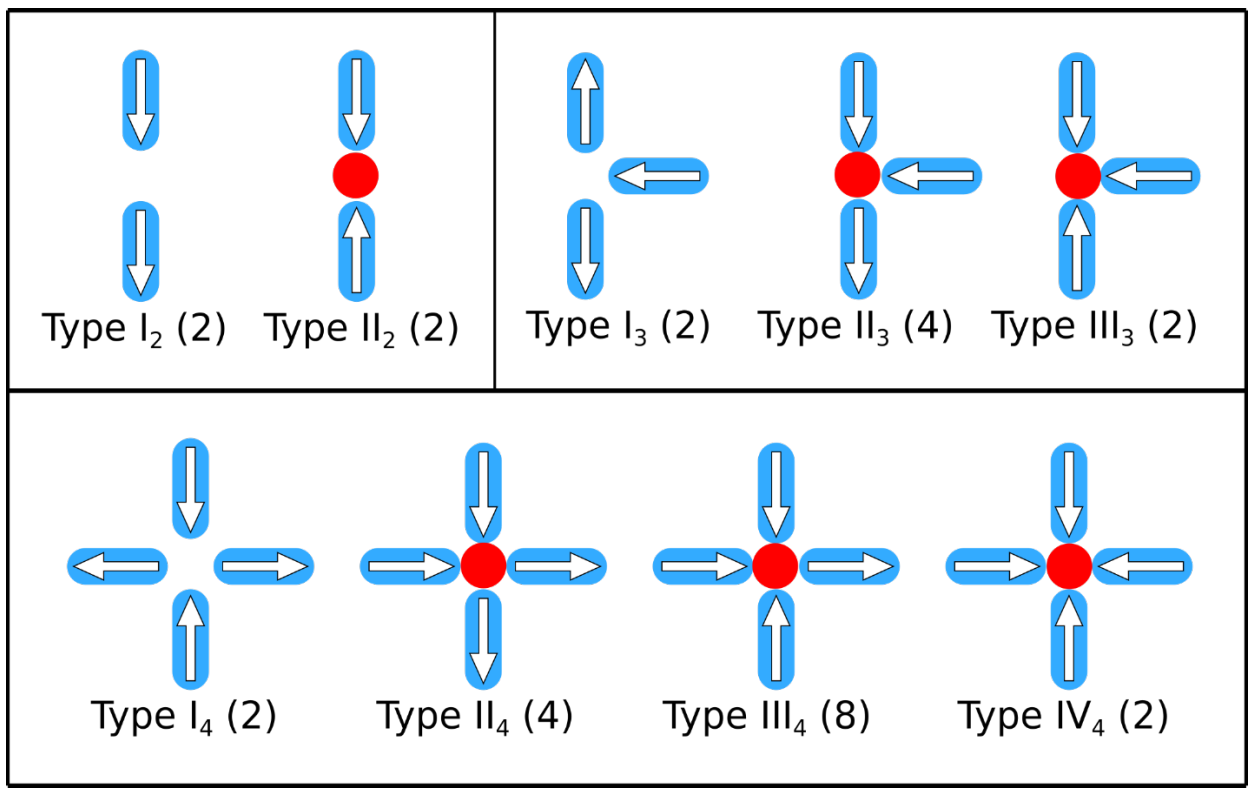

**Fig. S6. Vertex moment configurations for different vertex coordination numbers $z$ = 2, 3, and 4, arranged in order of increasing energy.** The arrows indicate moment direction and the red circles denote an excited state of the vertex (i.e., an unhappy vertex). The numbers in parentheses show the degeneracy for each given vertex type.

| SFI array | Length (nm) | Width (nm) | *Magnetostatic* energy along long-axis | *Magnetostatic* energy along short-axis | Energy Barrier (K) |
|---|---|---|---|---|---|
| 600nm | 482±1.5 | 189±1.5 | 7.42 | 18.33 | 79100 |
| 700nm | 477±4.9 | 186±3.8 | 7.21 | 18.15 | 79200 |
| 800nm | 477±1.2 | 179±1.5 | 7.01 | 18.16 | 80800 |

**Table S1. Magnetostatic energies of single islands.** Energies are giving with moment aligned with the easy-axis, perpendicular to the easy-axis, and with energy barrier calculated by micromagnetic simulation program MUMAX3. The energies are in unit of $10^{-19}$ J.

| SF array | $E(I_2)$ | $\Delta E_{II}^2$ | $E(I_3)$ | $\Delta E_{II}^3$ | $\Delta E_{III}^3$ | $E(I_4)$ | $\Delta E_{II}^4$ | $\Delta E_{III}^4$ | $\Delta E_{IV}^4$ |
|---|---|---|---|---|---|---|---|---|---|
| 600nm | 1.42 | 12.3 | 2.00 | 6.31 | 44.5 | 2.61 | 3.30 | 23.2 | 94.3 |
| 700nm | 1.41 | 5.58 | 2.07 | 3.29 | 20.3 | 2.73 | 3.60 | 11.7 | 41.7 |
| 800nm | 1.39 | 3.13 | 2.05 | 1.95 | 11.6 | 2.72 | 2.25 | 6.87 | 23.7 |

**Table S2. Magnetostatic energies for the ground state and the excess energy for all the excited states of $z$ = 2, 3, and 4 vertices.** The energies are calculated by micromagnetic simulation program MUMAX3. The energies of ground states are in unit of $10^{-18}$ J and the excess energies are in units of $10^{-20}$ J.